\documentclass[AMA,LATO1COL]{WileyNJD-v2}

\usepackage{float}
\usepackage{amssymb}
\usepackage{comment}
\usepackage{soul}
\usepackage{setspace}
\usepackage{graphicx}
\usepackage{subcaption}

\articletype{Article Type}%

\received{dd Month 2019}
\revised{dd Month 2019}
\accepted{dd Month 2019}
\let\OLDthebibliography\thebibliography
\renewcommand\thebibliography[1]{
  \OLDthebibliography{#1}
  \setlength{\parskip}{0.5pt}
  \setlength{\itemsep}{0.5pt plus 0.3ex}
}

\newboolean{showcomments}
\setboolean{showcomments}{true} 
\ifthenelse{\boolean{showcomments}}
{\newcommand{\nb}[2]{
  \fcolorbox{black}{yellow}{\bfseries\sffamily#1}
  {\sf$\blacktriangleright$\textit{#2}$\blacktriangleleft$}
 }
 
}
{\newcommand{\nb}[2]{}
 
}

\begin{document}



  \thispagestyle{empty}
      \newpage

    \setcounter{page}{1}
    \setcounter{section}{0}

\title{Blockchain and Cryptocurrencies: a Classification and Comparison of Architecture Drivers}

\author[1,2,6]{Martin Garriga*}
\author[1,2]{Stefano Dalla Palma}
\author[3]{Maxmiliano Arias}
\author[3]{\\Alan De Renzis}
\author[4]{Remo Pareschi}
\author[2,5]{Damian Andrew Tamburri} 

\authormark{M. Garriga \textsc{et al}}

\address[1]{\orgdiv{Department of Management}, \orgname{Tilburg School of Economics and Management}, \orgaddress{\state{Tilburg}, \country{The Netherlands}}}

\address[2]{\orgdiv{JADE Lab}, \orgname{Jheronimus Academy of Data Science}, \orgaddress{\state{Den Bosch}, \country{The Netherlands}}}

\address[3]{ \orgname{Fidtech}, \orgaddress{\state{Neuquen}, \country{Argentina}}}

\address[4]{\orgdiv{Department of Biosciences and Territory}, \orgname{University of Molise}, \orgaddress{\state{Molise}, \country{Italy}}}

\address[5]{\orgdiv{Technical Univ. of Eindhoven}, \orgaddress{\state{Eindhoven}, \country{The Netherlands}}}

\address[6]{ \orgname{CONICET, Faculty of Informatics, National University of Comahue}, \orgaddress{\state{Neuquen}, \country{Argentina}}}

\corres{*Corresponding author. \email{m.garriga@jads.nl}}


\abstract[Summary]{ 
Blockchain is a decentralized transaction and data management solution, the technological leap behind the success of Bitcoin and other cryptocurrencies. As the variety of existing blockchains and distributed ledgers continues to increase, adopters should focus on selecting the solution that best fits their needs and the requirements of their decentralized applications, rather than developing yet another blockchain from scratch. In this paper we present a conceptual framework to aid software architects, developers, and decision makers to adopt the right blockchain technology. The framework exposes the interrelation between technological decisions and architectural features, capturing the knowledge from existing academic literature, industrial products, technical forums/blogs, and experts' feedback. We empirically show the applicability of our framework by dissecting the platforms behind Bitcoin and other top 10 cryptocurrencies, aided by a focus group with researchers and industry practitioners. Then, we leverage the framework together with key notions of the Architectural Tradeoff Analysis Method (ATAM) to analyze four real-world blockchain case studies from industry and academia. Results shown that applying our framework leads to a deeper understanding of the architectural tradeoffs, allowing to assess technologies more objectively and select the one that best fit developers needs, ultimately cutting costs, reducing time-to-market and accelerating return on investment.}

\keywords{Blockchain, Framework, Cryptocurrencies, Software Architectures}

\maketitle

\section{Introduction}\label{sec:intro}

Blockchain is a decentralized ledger, concurrent transaction and data management solution, well-known for being the technology behind the success of Bitcoin cryptocurrency \cite{nakamoto2008bitcoin}.
Its main goal is to create a decentralized environment with no third-party control over transactions and data\cite{huumo2016current}.
This technology is now mainstream as it addresses transactions-management in an unprecedented way featuring complete trust-based, anonymous transactions processing. Gartner estimates the business value of blockchain technology around \$176 billion by 2025 and \$3.1 trillion by 2030\footnote{\url{https://www.gartner.com/doc/3627117/forecast-blockchain-business-value-worldwide}}. The prime reason is the already widespread adoption of blockchain in financial transactions and cross-border payments \cite{hileman2017global}. 

Even though the most popular blockchain implementation is Bitcoin, a myriad others are currently running or in advanced development. 
A number of important non-financial applications are emerging for different use cases that vary from logistic and supply chain management \cite{hackius2017blockchain, ahmed2017food}, 
to those concerning the development of scalable, efficient and high-impact decentralized solutions to social, economic and environmental innovation challenges, including energy, life science and health care \cite{mengelkamp2018blockchain, kewell2017blockchain,genesy2019, ribitzky2018pragmatic}.

This led to an exponential increase of blockchain adoption in recent years, initiated by the boom of ICOs (Initial Coin Offerings). Nowadays, the ICO's hype has passed\footnote{\url{https://www.gartner.com/smarterwithgartner/the-reality-of-blockchain/}} as government regulations are catching up\footnote{U.S. Securities and Exchange Commission -- \url{https://www.sec.gov/ICO}}. Thus, it makes no sense to "reinvent the wheel" by building a custom blockchain from scratch every time; but rather to leverage, and probably combine existing, battle-hardened solutions to support new applications.

However, blockchain is not a silver bullet, and every implementation present significant tradeoffs on certain usage scenarios. For example, in need of low latency or near real-time computations, as the blockchain introduces sensitive delays due to the computational cost related to the (purposely) heavy processing required. Also, when Privacy is required, as information on a blockchain is typically available to all participants~\cite{xu2017taxonomy}. Furthermore, throughput scalability is also an issue, as mainstream public blockchains can only handle on average 3-20 transactions per second\footnote{\url{https://blog.ethereum.org/2016/01/15/privacy-on-the-blockchain/}}, two orders of magnitude behind traditional payment services such as VISA. Last but not least, blockchain has its inherent complexity and many configurations and variants that should be weighted as well~\cite{xu2016blockchain}.

When designing decentralized applications upon blockchain, practitioners should systematically consider the key technical capabilities and configurations, and assess their interplay and tradeoffs on quality attributes of the application's architecture \cite{xu2017taxonomy}. Moreover, a common grounding for comparison is needed to determine the most appropriate blockchain or distributed ledger implementation (if any) for a given application\cite{Hintzman17scte}.

In this paper we introduce \textit{ChainMaster}: a conceptual framework to aid software practitioners and decision makers to assess and adopt the most suitable blockchain or distributed ledger technology\footnote{For the sake of simplicity we will use the word blockchain to refer to any distributed ledger implementation, except when explicitly clarified.} for their decentralized applications. Main architectural features and their interplay with technical decisions are disclosed. The framework captures the current state of the art as well as the state of practice, as it is built upon a literature review encompassing both \textit{white} literature (academic papers) and \textit{grey} literature (white/technical papers of industrial products, technical forums/blogs, etc.), and subsequent refinement with practitioners. 

Additionally, during framework construction we elicited feedback from experts through focus groups, questionnaires and semi-structured interviews, targeting twelve software practitioners and researchers from 5 different companies and 3 universities both in Europe and South America. Based on their insights we iteratively refined the framework and further developed a \textit{blockchain scorecard} that allows one to assess suitability of blockchain implementations for a given use case. Then we applied the scorecard alongside practitioners to assess most popular blockchains and distributed ledgers -- i.e., the ones behind Bitcoin and the following top 10 cryptocurrencies\footnote{According to their market cap. See: \url{https://coinmarketcap.com}} -- empowering developers to make informed decisions by moving from tacit to explicit knowledge on blockchain perils and opportunities.


Subsequently, we empirically validated our conceptual framework by using it in conjunction with key notions of the Architectural Tradeoff Analysis Method ATAM\cite{kazman1998architecture}  (namely, scenarios and utility trees), to assess four real-world blockchain case studies from industry (insurtech and artworks market domains) and academia (optimization and genomics research domains). Together with the developers of such applications, we assessed technologies more objectively and gained insights on the interplay of their own requirements and scenarios with blockchain architectural features. Ultimately, the process led to a selection (or at least an informed decision) of the most suitable technologies for their distributed applications.

As main findings, we unveil the key architectural features of blockchain (and other distributed ledgers) arising from the state of the art and state of practice, namely Cost, Consistency, Functionality \& Extensibility, Performance \& Scalability, Security, Decentralization and Privacy. Also, their interplay with the technical decisions behind blockchain implementations (transaction fees, consensus protocols, permission schemes, etc.) suggests that they cannot be addressed in isolation, but rather seen as a set of tradeoffs. For example, Scalability, Security and Decentralization cannot be achieved altogether, as they depend on conflicting requirements upon the permission schema and consensus protocols, among others. This holistic analysis allows to improve the quality of blockchain-based solutions, cutting costs and accelerating time to market and return on investment.

To summarize, the contributions of this paper are threefold. First, the conceptual framework capturing main architectural features and their interrelation with technological decisions on blockchain. Second, an analysis of the blockchain and distributed ledger solutions behind the top cryptocurrencies in the market. Finally, the empirical validation comprising the four case studies as example usage scenarios for our framework. 

The paper is organized as follows. Section~\ref{sec:background} provides background in blockchain and cryptocurrencies. Section~\ref{sec:related} discusses related work. Section~\ref{sec:framework} details the construction of our framework for assessment of blockchain technologies. Section~\ref{sec:assessment} assesses the top blockchain solutions by means of our framework. Section~\ref{sec:case-study} empirically validates the framework by applying it to four real-world case studies. Finally, Section~\ref{sec:conclusion} concludes the paper.

\section{Background} \label{sec:background}

A \emph{blockchain} is a distributed ledger, in the form of a totally ordered, back-linked list of blocks \cite{nakamoto2008bitcoin}.
Each block contains \textit{transactions} hashed into a binary tree (a \emph{merkle tree}), with the hash of the root (the genesis block) stored alongside. Each block also contains the previous block's hash, thus
guaranteeing integrity, replicability and determinism --- i.e., any node replicating all transactions starting from the first block will arrive to the same state as any other node \cite{buterin2014next}. This disables calling external APIs whose responses may change over time\footnote{\url{http://www.truthcoin.info/blog/contracts-oracles-sidechains/}}. Decentralization needs also  a consensus mechanism for trust creation and for agreeing on the next block to append. \emph{Cryptocurrencies} were the first intended application of blockchain technology, as digital currencies based on cryptography techniques and peer-to-peer networks\cite{garriga2018blockchain}. The first and most popular example is Bitcoin \cite{nakamoto2008bitcoin,alharby2017blockchain}.

One of the fundamental disruptions of blockchain technology is the questioning and then redefinition of digital \emph{trust}: trust emerges in a fully distributed way, without any node having to trust any single other member of the network. The only trust required comes from the consensus algorithm, where the majority of the nodes in the network should not colluding against the others in a coordinated manner \cite{xu2017taxonomy}.

\emph{Transactions} are the way to interact with the blockchain, in the form of data packages that store information --- e.g., monetary value for cryptocurrencies, executable application code, or results calling functions of \textit{decentralized applications} (\textit{dApps}). The integrity of a transaction is programmatically checked by the nodes using cryptography. A transaction signed by its sender, is propagated to the nodes connected to the blockchain network that validate and further propagate it, until the transaction reaches all nodes in the network \cite{xu2017taxonomy}. Transaction processing is charged with a \emph{transaction fee}, which is the reward for the load imposed to the nodes, and an incentive for nodes to stay honest \cite{nakamoto2008bitcoin,garriga2018blockchain}. 

Transactions are grouped in blocks of fixed size that are added to the existing chain in the process known as \emph{mining}. Nodes in the network aim to reach a \emph{consensus} regarding the next block to append by means of a \emph{Consensus Protocol}. Such protocol is the core of the blockchain, as it ultimately ensures decentralized governance, quorum, performance, authentication, integrity, non-repudiation and byzantine fault tolerance\cite{mattila2016blockchain}. The de-facto consensus protocol (Bitcoin and Ethereum) is Proof-of-Work \cite{back2002hashcash} (PoW). PoW miners compute a hash function that should be efficiently verifiable, but parameterisably expensive to compute, e.g., a hash with a given number of leading zeroes (\textit{nonce}). However, due to the vast energy consumption of PoW\footnote{\url{https://digiconomist.net/bitcoin-energy-consumption}}, and the centralization occurring in practice due to nodes colluding into mining pools,
newer blockchains implement greener but rather centralized algorithms, such as Proof-of-Stake \cite{kiayias2017ouroboros} (PoS), where miners obtain the right to mine blocks corresponding to their ownership stake of the token, 
and then Delegated Proof of Stake (DPoS). More recently, other blockchains implemented their own hybrid or ad-hoc protocols \cite{sankar2017survey}.

Bitcoin and other first generation blockchains were intented to support cryptocurrencies; as such they provided limited capability to support any business logic, apart from value transfer between accounts. The second generation (e.g., Ethereum \cite{buterin2014next}) was conceived as a general-purpose peer-to-peer programmable
infrastructure, supporting complex programs known as \emph{smart contracts} \cite{szabo1997formalizing,xu2017taxonomy}. Originally, smart contracts were intended as the digital equivalent of a paper contract: an agreement between parties with a set of promises that are legally enforceable \cite{szabo1997formalizing}. Nowadays, the definition of smart contracts extended to any general purpose computation that takes place on a blockchain or distributed ledger\cite{garriga2018blockchain}.

\section{Related Work} \label{sec:related}

There exist some incipient research work on deciding whether a blockchain or non-blockchain  (e.g., traditional/distributed databases) solution suits a given problem or domain. Lo et al.\cite{lo2017evaluating} developed a suitability evaluation framework which guides developers through a series of architecture-related questions -- encompassing multi-tenancy, trusted authorities, decentralization and more -- to decide between blockchain and conventional databases. In turn, W\"{u}st and Gervais\cite{wust2018blockchain} analyze the blockchain suitability problem accounting the permissionless/permissioned spectrum, and the properties behind it, namely verifiability, transparency, privacy, integrity, redundancy and trust. Their work also presents a series of questions in the form of a flow chart to drive the decision. However, we assume that the decision of adopting blockchain (or a distributed ledger) has been already made by the developers/organization, and therefore provide a means to discern among the myriad of different implementations in the market.

Yli-huumo et al. \cite{huumo2016current} extracted the technical challenges behind most of the blockchain-related papers to the moment:
security, usability, throughput, latency, size and bandwidth, wasted resources, versioning, hard forks, and multiple chains interplay\cite{swan2015blockchain}. Their work adds privacy, smart contracts, consensus protocols, new cryptocurrencies, botnets and trustworthiness as well. As future research directions, the authors highlight scalability constraints (the majority of blockchains can only process a few dozens or hundreds of transactions per second); other uses beyond cryptocurrency systems (i.e., dApps) and effectiveness of the proposed solutions (which is one of our contributions). In a similar direction, Alharby and Van Morsel \cite{alharby2017blockchain} focused on smart contracts, providing a systematic mapping in such field. They identified four groups of research efforts, according to the challenges they tackle: codifying, security, privacy and performance.

Scriber \cite{scriber2018framework} performed a literature review together with an evaluation of 23 blockchain implementation projects at early stages. The evaluation revealed 10 architectural characteristics that can drive the decision of whether a blockchain is suitable for a given problem: trust, immutability, transparency, identity, transactions, distribution, historical record, ecosystem and (in)efficiency. Although the characteristics may be relevant, from the 23 analyzed projects only four reached an advanced stage while the others stalled or failed for different reasons. Thus, the paper does not provide insights regarding which of the existing blockchains would be suitable for a given case.

Xu et al. \cite{xu2017taxonomy}, present a taxonomy of blockchain systems, their characteristics, and their impact on the design and assessment of software architectures. 
First, it identifies the fundamental properties of blockchain, namely: immutable data,  transparency, non-repudiation, data integrity, equal rights (of participants), and trust mechanisms. Other arising properties include data privacy, scalability, cost and performance. Afterwards, the work captures architectural design issues that impact such properties, namely: decentralization, support for client storage and computation, scope (public/consortium-community/private), data structure, consensus protocol and new side-chains. Then, it navigates a set of decisions in the form of a flowchart, that may affect those drivers when designing a blockchain from scratch. Differently to our work, the decisions correspond to the design and implementation of a new blockchain, rather than evaluating the adoption of an existing one.

\section{ChainMaster: A Conceptual Framework for Blockchain Architectures} \label{sec:framework}

In this section we present \textit{ChainMaster}, our conceptual framework for blockchain and related architectures\footnote{A preliminary, unpublished draft of the framework can be found at \cite{garriga2018blockchain}.}. Note that the main goal of the framework is to aid developers in the process of comparing, assessing, and ultimately selecting a blockchain or distributed ledger for their distributed applications. Framework construction is presented in Section~\ref{sec:fw-construction}.
Section~\ref{sub:chainmaster-arch-dimensions} presents the core of \textit{ChainMaster}: a set of architectural features, analogous to quality attributes, along with their relation with blockchain-oriented technical design decisions. Finally, Section \ref{sub:summary} provides a summary and a brief discussion on the interplay between the architectural features and technical decisions in Table~\ref{tab:arch-tech}. Such interplay and arising tradeoffs will be discussed in depth throughout the rest of the paper.

\subsection{Framework Construction}
\label{sec:fw-construction}

The first step towards framework construction is a Literature Review (LR), performed according to the key guidelines proposed by Kitchenham~\cite{Kitchenham07}. Although a systematic LR is out of the scope of this work, this helped us organize the process of finding and classifying relevant work. We searched for articles in the broad topic of blockchain architectures in different online databases, namely Scopus~\cite{scopus}, Science Direct~\cite{sciencedirect}, Wiley Online~\cite{wiley}, IEEExplore~\cite{ieeexplore}, Springer Link~\cite{springerlink} and ACM~\cite{acmonline}. The search strings (depicted as regular expression) were:

\vspace{-.2cm}
\begin{verbatim}
            blockchain & [architecture(s) | tradeoffs | benchmark | features | framework | applicability]
\end{verbatim}
\vspace{-.2cm}

We considered journal, conference and workshop articles (given the novelty of the topic) published up to 2018, the moment of performing our search. Then, we applied snowballing~\cite{wohlin2014guidelines}, by looking for relevant references in previously found articles. This allowed us to encompass in our study not only blockchain but also related distributed ledger technologies.

Then we applied Inclusion/Exclusion criteria to narrow the results set by including \textit{only} those that provide insights regarding architectural features on blockchain and their tradeoffs, in the form of SLRs, benchmarks, comparative frameworks, etc. Based on quality assessment control factors, we excluded at this point primary studies reporting new blockchain implementations, example applications or case studies. All in all, we ended up with $15$ selected academic papers. As the number of available and relevant studies demonstrated to be low, we extended the scope by including \textit{grey} literature
\cite{garousi2016grey}, intended as materials and research produced by organizations outside of the traditional commercial or academic publishing and distribution channels. Common \textit{grey} literature publication types include reports (annual, research, technical, project, etc.), working papers, government documents, white papers and blog posts. Even though the use of \textit{grey} literature is risky since there is often little or no scientific factual representation of data or analyses presented in \textit{grey} literature itself, its combination with traditional literature to determine the state of the art and practice around a topic is gaining a considerable interest in many fields, including software engineering\cite{garousi2016grey,soldani2018pains}. 

We circumscribed our \textit{grey} literature review to the top 11 cryptocurrencies -- Bitcoin plus 10 as of August, 2018, and the list is mostly stable -- exploiting general Web search engines with the following search strings:

\vspace{-.2cm}
\begin{verbatim}
    [bitcoin | ethereum | ripple | bitcoin cash | eos | litecoin | stellar | cardano | ada | iota | tether | tron ] 
                        & [white paper | technical paper | benchmark | features ]
\end{verbatim}
\vspace{-.2cm}

Then, as done before for the academic literature we applied snowballing into the references and links within the results, and applied inclusion/exclusion criteria and quality assessment control factors. The goal here is to circumscribe the results to white/technical papers (and their accompanying material) of the topmost cryptocurrencies, and then other sources such as blog posts analyzing technical decisions and features, or performing benchmarks and testing upon such currencies. The dataset of \textit{grey} literature is then comprised of $25$ entries. Thus, the whole dataset was composed of $40$ entries: $15$ and $25$ from the white and \textit{grey} literature reviews respectively\footnote{A summary of the dataset is available at: \url{https://tinyurl.com/y7gxj53e}}. 

Following, we analyzed the selected dataset by applying notions of Grounded-Theory\cite{glaser2017discovery}. In this approach, 
a series of systematic steps allow theory to emerge from the data (hence, "grounded"). In our case this implies iterative coding (i.e., tagging) the content of the sources in our dataset that referred in any way to architectural features -- e.g., definitions, discussions, tradeoffs, etc. After reaching a consensus on the features by the two authors performing the coding, such initial list was delivered, together with the sources dataset, to a focus group of five experts: two researchers from academia, two developers of blockchain-based applications, and one blockchain entrepreneur from one of the top-10 blockchains. These experts performed an inter-rater reliability assessment\cite{hallgren2012computing} based on their own expertise and supported by the literature. As the participants have expertise on the blockchain field, they were given only general guidelines: to assess the set of architectural features and suggest to add, remove, group, decompose them, or point out any other possible correlation. After the first revision round, the k-$\alpha$ score regarding the consensus on the list of features was of $0.7$, which is high but can be improved. After the refinement of the features, we reached a k-$\alpha$ score of 0.875. All in all, the final list of features stemming from the whole process of Grounded Theory plus experts feedback is summarized in Table \ref{tab:arch-tech}, and discussed below.


\subsection{ChainMaster's Architectural Dimensions}
\label{sub:chainmaster-arch-dimensions}

Our analysis reveals \textit{seven} key architecture features, as distilled from the literature by means of Grounded theory. The seven features are: (1) Cost; (2) Consistency; (3) Functionality \& Functional Extensibility; (4), Performance \& Scalability; (5) Security; (6) Decentralization; and (7) Privacy. The following sections define the dimensions and outline the technical (design and implementation) decisions that affect them. These dimensions stem mainly from blockchain platforms, although applicable for other distributed ledger technologies. Throughout the section we refer to blockchains except when explicitly stated otherwise.


\subsubsection{Cost}
Adopting a blockchain is theoretically free, but at least three aspects impose a cost for using the network: a variable cost for running transactions, composed by the \emph{transaction fee} \cite{buterin2014next} and \emph{incentives} for processing transactions; and a minimal fixed cost to deploy applications as \textit{smart contracts}.

The default approach is to have purely voluntary fees with dynamic minimums \cite{nakamoto2008bitcoin}. However, this approach can become prohibitively expensive when the network is congested: for example, transaction fees in Bitcoin have raised up to 40 USD by December 2017\footnote{\url{https://bitcoinfees.info/}} during peaks of workload. On the other end, implementations such as Tron\cite{tron} foster minimal transaction fees to prevent malicious users to perform DDoS attacks for free. 

\textcolor{red}{
Ethereum assigns a deterministic cost, named gas, to EVM (Ethereum Virtual Machine) operations composing a transaction. The sum of all operations determines the overall cost of the transaction. Then, the user assigns to each gas unit a price, named gas price, thus determining the fee the user has to pay: higher the fee, sooner the transaction gets mined.}
This is not so different from paying a fee per byte of transaction size as suggested in Bitcoin. During a period of high congestion, both can cause high fees races to gain the space available in blocks. Yet another approach is to impose a non-monetary fee for running transactions. This is indeed an innovative feature brought by other non-blockchain distributed ledgers such as IOTA\cite{iota}, where every node sending a transaction is required to validate two other transactions, which assures enough processing power\cite{popov17tangle}. 

Possible values for the \emph{transaction fee }are: Minimal (only to avoid DDoS attacks), per transaction and per instruction. Possible values for the \emph{incentive} are: Big (bitcoin-alike), and small (equivalent to a fee).

\subsubsection{Consistency}

Different strategies have been used to confirm that a transaction
is securely appended to the blockchain, mostly to wait for a certain number of blocks (e.g., 6 for Bitcoin, 12 for Ethereum) to have been generated after the transaction is considered consistent into the blockchain. 
However, none of the current implementations can ensure strong \emph{consistency}, but only eventual consistency that tends to (but never reaches) 100\% as more time passes and more blocks are confirmed. This implies a tradeoff between safety and time\cite{decker2016bitcoin}.

Alternative implementations of the distributed ledger other than blockchain, for example a Directed Acyclic Graph\cite{dag} (DAG) as in IOTA, can drastically reduce the time to confirmation as they do not rely on blocks with multiple transactions, but in transactions that are propagated independently in a matter of seconds\cite{garriga2018blockchain}. Thus, consistency is a function of the \emph{time to confirmation} (i.e., the number of blocks after which one can consider a transaction securely appended to the blockchain), which in turn depends on the \emph{block production rate} (BPR, the amount of time required to mine a block), configured for each implementation at design time. Possible values for \emph{time to confirmation} are: seconds, minutes
and hours. Possible values for \emph{block production rate} are: 10 minutes or more, 1 to 10 minutes, and seconds.

\subsubsection{Functionality and Extensibility}

According to the Bitcoin whitepaper, blockchain's main intent was to support a decentralized and electronic payment system based on cryptography instead of trust~\cite{nakamoto2008bitcoin}. Rapidly, the idea of applying this technology to other concepts such as decentralized applications (\textit{dApps}) emerged, e.g., for name registration and tokens for corporate use \cite{buterin2014next}. 
Thus, the feature Functionality and Extensibility refers to the ability of supporting such a complex behavior by a blockchain implementation: either as a built-in \textit{functionality} or through \textit{extension} mechanisms allowing developers to craft applications beyond the original intent of the platform. One may note that basic functionality such as transactions and encryption is assumed here\cite{garriga2018blockchain}.

This can be achieved through supporting smart contracts. Some implementations support turing-complete smart contracts using ad-hoc languages (Solidity in Ethereum, Plutus in Cardano), while others support traditional languages (such as C++ in EOS) that are then compiled/transpiled to a bytecode. However, the latter are not fully supported yet in any of the existing blockchain implementation.

A latter point is \emph{interchain communication}, allowing multiple parallel blockchains to interoperate retaining their security properties \cite{kwon2016cosmos}. 
\textcolor{red}{
Some blockchain implementations provide native support for interchain communication. For example,  EOS\cite{eos18white} uses Merkle Proofs for light client validation (so clients do not need to process all transactions), while Cosmos provides native inter-blockchain communication (IBC) on top of Tendermint Core\cite{kwon2016cosmos}. Sometimes, adapters may be needed to bridge initially different implementations, such as Peg Zones in Cosmos~\cite{kwon2016cosmos}, which allow to communicate with other (BTC, ETH) existing chains. } Possible values for \emph{smart contracts} are: Yes (specifying the language(s)), No, and Very Limited. Possible values for \emph{Interchain communication} are: Yes, No.

\subsubsection{Performance and Scalability}

Decoupling \emph{performance} (latency and throughput) and \emph{scalability} (with the number of nodes and clients in the system) is not entirely possible \cite{vukolic2015quest}, thus we will group them together for analysis. Those became the bottleneck for the most popular blockchain implementations, such as Bitcoin (consensus latency of about an hour) and Ethereum. Additionally, PoW-based networks use a lot of power, equivalent to a small country such as Austria \cite{vukolic2015quest}. 
The introduction of different permission schemes could also affect performance: if only few nodes or a single organization can validate transactions (e.g., Ripple and its consensus protocol)\cite{garriga2018blockchain}, certainly this can offer better performance compared to permission-less blockchains. In short, centralized validation mechanisms are simpler, but they may also compromise security, as discussed in next section.



Scalability, in turn, refers to the ability to maintain performance indicators when serving more users and transactions, limited by: (i) the size of the data on blockchain, (ii) the transaction processing rate, and (iii) the latency of data transmission. Roughly speaking, PoW offers good scalability with poor performance, whereas other protocols offer good performance for small numbers of replicas, with limited scalability. Given seemingly inherent trade-offs between the number of nodes and performance, it is not clear today what the optimal blockchain solution is, for the majority of use cases in which the number of nodes ranges from a few tens to a few thousands.

Conclusively, performance and scalability are affected by the \emph{transactions per second} (TPS), \emph{block production rate, consensus protocol}, the \textit{permission schema}, and certain \emph{technology enhancements} on top of the blockchain. Possible values for \emph{transactions per second} are: less than 100, between 100 and 1000; and 1000 or more. Possible values for the \emph{Consensus Protocol} are: Proof-of-Work (PoW), Proof-of-Stake (PoS), Delegated Proof-of-Stake (DPoS), and other (specify). Possible values for \textit{permission schema} are permissioned and permissionless.
Possible values for technology enhancements are: Merkle/Patricia trees, Segregated Witness (segwit), data sharding, parallel execution of transactions, GHOST (Greedy Heaviest Observed Sub-Tree), Lightning Network, and other (specify).

\subsubsection{Security}

One of the main features of blockchain is that its public ledger cannot be modified or deleted after the data has been approved by all nodes, providing data integrity and security characteristics \cite{huumo2016current}.
Security issues mean bugs or vulnerabilities that an adversary might
utilize to launch an attack\cite{garriga2018blockchain}. Currently, the most secure implementations are PoW-based. Even though, they have a possibility of a 51\% attack, where a single entity would have full control of the majority of the network's mining hash-rate and would be able to manipulate it. Furthermore, studies show that \textcolor{red}{selfish mining pools exceeding 33\% of the hash power will always be able to collect mining rewards that exceed its proportion of mining power, even if it loses every single block race in the network~\cite{eyal2014majority}.}

Alternative consensus protocols such as PoS and DPoS may provide better
performance and/or scalability, but they imply a tradeoff w.r.t. security: most tolerate up to 1/3 (33\%) of malicious nodes. Other algorithms may improve security up to 2/3 malicious nodes, but they impose additional restrictions such as requiring nodes to know each
other. Security is thus affected by the following technical decisions:
\emph{Fault tolerance} (possible values are 2/3 attack, 1/2 attack,
1/3 attack, and other); \emph{consensus protocol}, and \textit{permission schema}.

\subsubsection{Decentralization}
Theoretically, blockchain does not rely on any centralized node or authority, allowing data to be recorded, stored and updated in a distributed fashion. However, some blockchains introduce certain degree of centralization. In case of public, permission-less blockchain, no centralized authority or party has more power than the rest (Bitcoin), and everyone has the right to validate a transaction \cite{vukolic2015quest}.

In the case of consortium, permissioned blockchain, only few nodes are given certain privileges over validation -- e.g, in DPoS-based
ones such as EOS, or in alternative ledgers such as Hyperledger. A fully private blockchain has a centralized structure with the power to take decisions and control the validation process (e.g., Ripple and the Ripple Consensus Protocol)\cite{garriga2018blockchain}. Permissioned blockchains are faster, more energy efficient and easily implementable compared to permission-less blockchains, but introduce certain degree of centralization. 

\textcolor{red}{The effects of incentives on decentralization are quite subtle, as they intertwine with the adopted consensus algorithm. The more attractive the incentives, the more nodes are created and hence the more opportunities emerge for decentralization, even if the consensus protocol is inherently centralized. From this point of view, one can distinguish between \textit{de jure} and \textit{de facto} centralization. DPoS is centralized \textit{de jure}, while PoW tends to be centralized \textit{de facto}, through the aggregation of  miners into pools~\cite{li2020comparison}.} All in all, decentralization  is affected by the \textcolor{red}{\emph{incentives}}, \emph{consensus protocol}, the \textit{permission schema} and the \emph{ledger implementation}.

\subsubsection{Privacy}
Most blockchains are designed to protect transaction integrity, but they do not consider transaction privacy. A blockchain is said to have transaction privacy when (1) transactions cannot be linked from one to another, and (2) the transaction content is known only to its participants. Bitcoin, for example, does not guarantee anonymity but pseudonymity, in which transactions can still be tracked to their sender/receiver wallets (even though the wallet owners may remain unknown behind their pseudonym).

In public settings, all transactions and users' balances are publicly available to be viewed. The privacy setting is limited since there are no privileged users, and every participant can join the network to access all the information on the blockchain\cite{lo2017evaluating}.
This lack of privacy could hinder adoption as many people consider financial transactions (e.g., stock trading) as confidential information\cite{alharby2017blockchain}. 

In private settings, complete transparency of transaction
history may not be a problem\cite{wust2018blockchain}. Either transparency is desirable for the applications, such as financial auditing, or it is straightforward to add an access control layer to protect the blockchain data\cite{dinh2018untangling}.
Privacy is affected by the \textit{consensus protocol}, as it defines what needs to be seen by the different nodes/miners (e.g., Zerocoin implements a zero-knowledge proof scheme\cite{miers2013zerocoin}); and by the \textit{enhanced technology} that could provide extra obfuscation on top of a blockchain implementation, e.g. Hawk\cite{kosba2016hawk}, a tool to write privacy-preserving generic smart contracts that are then compiled to a target blockchain obfuscating sensible data.


\subsection{Summary and discussion}
\label{sub:summary}

Table~\ref{tab:arch-tech} summarizes the interplay between technical decisions and architectural features stemming from the Grounded-Theory process described in Section~\ref{sub:chainmaster-arch-dimensions}. Additionally, we can foresee the relationships and tradeoffs among the different architectural features, that call for further analysis.

Initially, Cost and Functionality features seem to be somewhat isolated from the other features, as they are governed by particular technical decisions: fees and incentives do not affect nor depend on other parameters; and the same goes for the support of smart contracts. Consistency is another rather isolated feature, although it has an interplay with Performance: as a higher block production rate could increase the latter, but moving towards eventual consistency (i.e., achieved at an arbitrary point in the future) rather than strong consistency\cite{decker2016bitcoin}.

Interesting tradeoffs lie in the triads formed by Performance (\& Scalability) -- Decentralization -- Security/Privacy, as they are governed by conflicting technical decisions. For example, increasing privacy by obfuscation allows to keep decentralization but at expenses of performance; while increasing privacy by means of trusted authorities mediating the transactions allows to keep performance but reducing decentralization. The same goes for the triad formed with security, a grounded confirmation of the scalability trilemma\cite{buterin2017trilemma}, where blockchain systems can only at most have two of the following properties: decentralization, avoiding a central authority/point of failure; performance/scalability, allowing a fast processing of transactions; and security, hardening the blockchain against a possible attack. We will further elaborate on these tradeoffs in the following sections, which leverage \emph{ChainMaster} for the analysis of top blockchain implementations.


\begin{table}[ht]
\caption{Summary of technological decisions affecting architecture features according to \textit{Chainmaster} framework, as discussed through Section~\ref{sub:chainmaster-arch-dimensions}. \label{tab:arch-tech}}
\centering%
\footnotesize
\begin{tabular}{l|ccccccc}
\toprule
\multirow{2}{*}{Technical decison} & \multicolumn{7}{c}{Architecture features} \\ \cline{2-8} 
                                   & Costs  & Consistency  & Functionality  & Performance & Security & Decentralization & Privacy  \\ \toprule
Fees                               & x  &   &   &   &   &  & \\ \midrule
Incentives                         & x  &   &   &   &   & \textcolor{red}{x} & \\ \midrule
Confirmation time                  &   & x  &   &   &   &  &\\ \midrule
Block production rate              &   & x  &   & x  &   & & \\ \midrule
Smart contracts                    &   &   & x  &   &   &  &\\ \midrule
Inter-chain                        &   &   & x  &   &   &  &\\ \midrule
Consensus                          &   &   &   & x  & x  & \textcolor{red}{x} & x\\ \midrule
Technology                         &   &   &   & x  &   & x & x \\ \midrule
Fault tolerance                    &   &   &   &   & x  &  &\\ \midrule
Ledger-Type                             &   &   &   &   & x  & x &\\ \midrule
TPS                                &   &   &   & x  &   &  & \\ \midrule
Permission schema                  &   &   &   & x  & x & x & x\\ \bottomrule
\end{tabular}
\end{table}

\section{Assessment of Blockchain Solutions} \label{sec:assessment}


In this section we present practical usage of \textit{ChainMaster} as a framework to assess blockchain solutions. 
First, Section~\ref{sec:assessment-tech} assesses the technical decisions of the blockchains and distributed ledgers underlying the most popular cryptocurrencies, under the light of \textit{ChainMaster}. Subsequently, Section~\ref{sub:blockchain-scorecard} arranges the architectural features in \textit{ChainMaster} as a \textit{blockchain scorecard} that allows practitioners to assess suitability of blockchain implementations when developing \textit{dApps}. The section follows-up with a focus-group featuring practitioners from industry and academia, who assessed the blockchains and ledgers behind the topmost cryptocurrencies using \textit{Chainmaster}, and discusses the main findings. 

\subsection{Blockchain Technological Decisions}
\label{sec:assessment-tech}

We analyzed the technological decisions in the blockchains and distributed ledgers behind the most popular cryptocurrencies -- according to their market cap\footnote{Available online at: \url{http://coinmarketcap.com}. Accessed on July 2018}. Even though we acknowledge other possible ways to filter blockchain implementations included in our study, to date no other use of blockchain has caused the same level of attention, investment and development other than cryptocurrencies. Moreover, other possible ways to measure popularity -- e.g., transaction volume \textemdash{} usually converge to a similar ranking at a given point in time \cite{hileman2017global}. 
Furthermore, the popularity of the underlying platform may be the key enabler for the success of a dApp, as demonstrated by the myriad of dApps in Ethereum\footnote{\url{https://github.com/avadhootkulkarni/UltimateICOCalendar}} and the limited offer in others. Needless to say, one can use our framework to assess other blockchains or distributed ledgers.

All in all, the analysis  is summarized in Table \ref{tab:qualitative-analysis}. The technological decisions were extracted from the Grounded-Theory presented in Section~\ref{sec:framework} -- i.e., those that affect the architecture features or quality attributes of the blockchain implementation. As for obtaining the values that fulfill the table, we relied on the information presented in the different white and technical papers, given they were demonstrated or empirically tested. Noticeably, some of the information was either obscure, ambiguous or non tested. For example, with regards to Transactions per Second (TPS), several implementations claimed to support thousands of transactions per second (e.g., EOS), but the calculations are the theoretical corollary of the algorithms used/proposed, and these numbers were not empirically tested whatsoever. \textcolor{red}{Regarding fault tolerance, even though some algorithms as PoW claim to support up to 50\% malicious nodes, selfish mining (in the presence of a single selfish miner) and similar block withholding attacks can succeed when 33\%+1 of the hash power is malicious\cite{eyal2014majority}.} 

Finally, in certain cases the main nets are not yet deployed and/or available for use. In such cases, when certain claims or values raised our concerns, we acquainted information from other backing studies in the (\textit{grey}) literature, performing benchmarks or empirical tests over the networks.  

\begin{table*}[ht]
\caption{Technology-based comparison of blockchains and distributed ledgers behind the most popular cryptocurrencies.\label{tab:qualitative-analysis}}

\resizebox{\linewidth}{!}{%
\begin{tabular}{lllllll}
\toprule
 & \textbf{Bitcoin}
\textbf{(BTC)} & \textbf{Ethereum} \textbf{ (ETH)} & \textbf{Ripple} \textbf{ (XRP)}\textsuperscript{$*$} & \textbf{Bitcoin Cash} \textbf{ (BCH)} & \textbf{EOS} \textbf{(EOS)} & \textbf{Litecoin} \textbf{(LTC)} \\
\toprule
\textbf{Purpose} & {Currency} & {Platform} & {Fintech} & {Currency} & {Platform} & {Currency} \\
\midrule 
\textbf{Fees} & {Per transaction} & {\textcolor{red}{Configurable GasPrice Per Instruction}} & {Minimal} & {Per Transaction} & {Per instruction} & {Per transaction}\\
\midrule 
\textbf{Incentive} & {6.25 \textcolor{red}{BTC/MinedBlock}\textsuperscript{$\dagger$}} & \textcolor{red}{{2.0 ETH/MinedBlock + Uncles}} & {N/A} & {6.25 \textcolor{red}{BCH/MinedBlock}\textsuperscript{$\dagger$}} & {Configurable (by BPs)} & {12.5 \textcolor{red}{LTC/MinedBlock}\textsuperscript{$\dagger$}}\\
\midrule 
\textbf{Block production rate} & {10 min} & {10-19 sec} & {Not specified} & {10 min} & {0.5 sec} & {2.5 min}\\
\midrule 
\textbf{Confirmation time} & {60 min (6 blocks)} & {1 min} & {seconds} & {60 min} & {1 second} & {20 min (6 blocks)}\\
\midrule 
\textbf{TPS} & {7} & {15} & {1500} & {23} & {1000+} & {50} \\
\midrule 
\textbf{Smart contracts} & {Very Limited} & {Yes} & {Very Limited} & {Very Limited} & {Yes} & {Very Limited} \\
\midrule 
\textbf{dApps} & {No} & {Yes} & {No} & {No} & {Yes} & {No} \\
\midrule 
\textbf{Languages} & {C++, Script} & {Solidity} & {Any} & {Script} & {Any} & {Script}\\
\midrule 
\textbf{Interchain} & {No} & {No} & {No} & {No} & {Yes} & {No}\\
\midrule 
\textbf{Consensus algorithm} & {PoW (sha-256)} & {PoW (ethash)} & {PoC (Ripple CP)} & {PoW (sha-256)} & {Delegated PoS } & {PoW (scrypt)}\\
\midrule 
\textbf{Enhanced technology} & {Merkle Trees, Segwit} & {Patricia Trees, Sharding} & {---} & {Merkle Trees} & {Merkle Proofs, Segwit} & {Merkle Trees, Segwit}\\
\midrule 
\textbf{Fault Tolerance} & {51\% CPU attack} & {51\% ether attack} & {20\% malus nodes} & {51\% CPU attack} & {33\% malus Block Producers} & {51\% CPU Attack}\\
\midrule 
\textbf{Ledger} & {Blockchain} & {Blockchain} & {Interledger Protocol} & {Blockchain} & {Blockchain} & {Blockchain}\\
\midrule 
\textbf{Permission schema} & {Permissionless} & {Permissionless} & {Permissioned} & {Permissionless} & {Permissioned} & {Permissionless}\\
\bottomrule
\multicolumn{5}{l}{\textsuperscript{$*$}\footnotesize{Distributed Ledger.} } & \\
\multicolumn{5}{l}{\textsuperscript{$\dagger$}\footnotesize{\textcolor{red}{At the time of writing, halved every 4 years. Last halvings in 2019 (LTC) and 2020 (BTC, BCH) }} } & 
\\\\

\toprule
& \textbf{Stellar} \textbf{ (XLM)} & \textbf{Cardano} \textbf{ (ADA)} & \textbf{Iota} \textbf{ (MIOTA)}\textsuperscript{$*$} & \textbf{Tether}
\textbf{ (USDT)} & \textbf{Tron} \textbf{ (TRX)} & \\
\toprule
\textbf{Purpose} & {Fintech} & {Platform} & {Other (IoT)} & {Currency} & {Media/Social} & \\
\midrule
\textbf{Fees} & {Per operation} & {Per transaction size} & {Approve 2 transactions\textsuperscript{$\dagger$}} & {20 USD} & {Minimal} & \\
\midrule
\textbf{Incentive} & {N/A} & {Availability based} & {N/A} & {No} & {Configurable by Super Representatives (SRs)}\\
\midrule
\textbf{Block production rate} & {5 seconds} & {20 sec configurable} & {N/A} & {10 min} & {15 sec}\\
\midrule
\textbf{Confirmation time} & {30 seconds} & {1 min} & {2 min (scalable)} & {60 min (6 blocks)} & {1 min} \\
\midrule
\textbf{TPS} & {1000+} & {250} & {1000+} & {7} & {1000+}\\
\midrule
\textbf{Smart contracts} & {Very Limited} & {Yes} & {No} & {No} & {Yes}\\
\midrule
\textbf{dApps} & {No} & {Yes} & {Yes, IoT level} & {Very Limited} & {Yes}\\
\midrule
\textbf{Languages} & {Any} & {Functional} & {Any} & {Ruby} & {Java}\\
\midrule
\textbf{Interchain} & {No} & {Yes} & {Yes} & {No} & {Yes}\\
\midrule
\textbf{Consensus algorithm} & {Federated BFTStellar CP} & {PoS (Ourboros)} & {Markov Chain (MCMC)} & {PoR (Omni)} & {PoS (Tendermint)}\\
\midrule
\textbf{Enhanced technology} & {\textendash{}} & {Sharding} & {Tangle for IoT Scale} & {Merkle trees, Segwit} & {Graph database}\\
\midrule
\textbf{Fault tolerance} & {Asymptotic Security} & {Asymptotic Security} & {51\% CPU attack} & {51\% CPU Attack} & {33\% Byzantine Failure}\\
\midrule
\textbf{Ledger} & {Blockchain} & {Blockchain} & {Directed Acyclic Graph} & {Blockchain} & {Blockchain}\\
\midrule 
\textbf{Permission schema} & {Permissioned} & {Permissioned} & {Permissioned} & {Permissioned} & {Permissioned}\\
\bottomrule
\multicolumn{5}{l}{\textsuperscript{$*$}\footnotesize{Distributed Ledger (Directed Acyclic Graph).} } & \\
\multicolumn{5}{l}{\textsuperscript{$\dagger$}\footnotesize{Instead of paying a fee in IOTA, a node submitting a transaction must validate two other transactions.} } & \\
\end{tabular}
}%
\end{table*}

\begin{table*}[ht]
\centering
\footnotesize
\caption{Quantitative assessment of blockchain solutions behind top cryptocurrencies, according to experts' feedback based on architectural decisions, technological features, and current status of the corresponding networks.}
\label{tab:quantitative-scores}
\begin{tabular}{lcccccccccccll}
\toprule 
 & {BTC} & {ETH} & {XRP} & {BCH} & {EOS} & {XLM} & {LTC} & {ADA} & {USDT} & {MIOTA} & {TRX} & & \tabularnewline
\cline{1-12}
\emph{Popularity} & \emph{1} & \emph{2} & \emph{3} & \emph{4} & \emph{5} & \emph{6} & \emph{7} & \emph{8} & \emph{9} & \emph{10} & \emph{11} & \textbf{Mean} & \textbf{Median} \\
\toprule 
{Cost} & {1.3} & {2.0} & {4.7} & {1.7} & \textbf{{5.0}} & {4.7} & {2.7} & {4.3} & \textbf{{5.0}} & \textbf{{5.0}} & \textbf{{5.0}} & \emph{3.8} & \emph{4.7} \\
\midrule 
{Consistency} & {1.3} & {2.3} & {4.3} & {1.3} & \textbf{{5.0}} & {4.0} & {2.0} & {3.7} & {1.0} & {4.7} & {4.0} & \emph{3.1} & \emph{3.7}\\
\midrule 
{Functionality} & {2.0} & \textbf{{5.0}} & {1.3} & {2.0} & \textbf{{5.0}} & {1.3} & {2.0} & {4.3} & {2.0} & {3.7} & \textbf{{5.0}} & \emph{3.2} & \emph{2.0} \\
\midrule 
{Performance} & {1.3} & {1.7} & {4.3} & {2.0} & {4.7} & {4.0} & {2.3} & {3.0} & {1.0} & \textbf{{5.0}} & {4.7} & \emph{3.1}  & \emph{3.0} \\
\midrule 
{Security} & \textbf{{4.0}} & \textbf{{4.0}} & {2.3} & \textbf{{4.0}} & {3.3} & \textbf{{4.0}} & \textbf{{4.0}} & \textbf{{4.0}} & {3.3} & {3.7} & {3.3} & \emph{3.6} & \emph{4.0}\\
\midrule 
{Decentralization} & \textbf{{5.0}} & {3.3} & {1.0} & {4.3} & {2.7} & {2.3} & {3.7} & {3.3} & {1.3} & {2.3} & {3.3} & \emph{3.0} & \emph{3.3}\\
\midrule
{Privacy} & 2.0 & 2.0 & 4.0 & 2.0 & 2.0 & 2.0 & 2.0 & 3.0 & 4.0 & 2.6 & \textbf{4.3} & \emph{2.7} & \emph{2.0} \\
\bottomrule
\end{tabular}
\end{table*}

\subsection{Blockchain Scorecard and Practitioners Assessment}
\label{sub:blockchain-scorecard}

Based on the identified architectural and technological aspects, we
conducted a focus group through semi-structured interviews
with twelve blockchain practitioners and researchers from five different companies and three universities both in Europe and South America, in order to come up with a numerical assessment
of the top blockchain solutions. The interviews were driven by a questionnaire\footnote{The questionnaire is available online at: \url{https://goo.gl/forms/8mE1mdJ55VJRJw2E3}}, where experts assigned scores for each architectural feature to the different blockchain implementations (from \emph{very low} to \emph{very high}), according to their perceptions on technological features, and current status of the corresponding networks. Answers were then fuzzified into a numerical scale from 1 to 5. For example,
if an expert considers that \emph{Bitcoin} has a very high \emph{Cost},
then in the questionnaire she marks the corresponding cell as "very
high". 

Participants filled the \textit{ChainMaster}'s scorecards as described, and declared their confidence (from \textquotedbl low\textquotedbl{} to \textquotedbl very high\textquotedbl). Both the confidence values and the scores were then defuzzified using a triangular membership function~\cite{pedrycz94triangular} and combined on a weighted average scheme. The triangular function allows one to map and normalize the linguistic values to a numerical scale, in the range {[}$1,5${]} for the scores and {[}$0,1${]} for the confidence. 

The result of the process is a normalized weight matrix, which numerically represent the scores for each feature on each blockchain implementation as values in the interval {[}$1, 5${]}, as shown in Table \ref{tab:quantitative-scores}. Although we acknowledge that the scores arising from a focus group with a reduced number of experts may be opinionated and not directly generalizable, the reflection exercise provides interesting insights on the promises and perils of each blockchain regarding the architectural features in \textit{ChainMaster}, and illustrate how to apply the framework. The scorecard allows software engineers and architects to assess existing blockchains to implement their dApps, as follows. First, they provide an importance score (from \emph{very low} to \emph{very high}) for each feature in the context of their dApp. Then such score is fuzzified and weighted with the score for each blockchain solution and feature in the scorecard. Finally the overall sum for each blockchain alternative represents its suitability in the context of the dApp under development. In case two or more alternatives attain the same overall sum, the decision makers can consider the scores for the most important factors in the context of their dApps -- e.g., if Security is the main concern, the selection should go towards the solution with the highest Security score. Furthermore, Section~\ref{sec:case-study} showcases four case studies that applied ChainMaster's rationale to assess blockchain alternatives.

There may be other factors that help untie between two alternatives: expertise of the team with a given language, community support, related projects on the same blockchain, open-source preferences, etc.  As non architectural factors, those are outside the scope of our framework. Finally, ChainMaster is not intended to be the single source of truth when choosing a blockchain solution, but to gain knowledge about the existing ones and insights on the \textit{dApp} requirements beforehand. The final decision is in hands of the software architect(s).

From the Cost perspective, the lower scores correspond to the higher transaction fees, as in Bitcoin.  As other performance parameters may be subject to upper bounds by design (e.g., TPS, block production rate and confirmation time), the only response to network congestion is to raise transaction fees. The higher scores correspond to fee-free or minimal-fee (just to avoid DDoS attacks) setups.


From the Consistency perspective, the lower scores are for those platforms that forked/copied the Bitcoin scheme of one block every 10 minutes and confirmation after 6 blocks (60 minutes). This is the case of Bitcoin itself, Bitcoin Cash (fork), Litecoin and Tether. The higher scores depict blockchains whose design allow for fast consistency by reducing confirmation time and increasing block production rate thanks to alternative consensus algorithms (DPoS in EOS) or ledger implementations (a DAG instead of a blockchain in IOTA). 


Functionality and Extensibility refer mainly to the support of (turing-complete) smart contracts. Low values imply the lack of smart contracts support, as the main goal of such implementations is to support cryptocurrency transactions (Bitcoin, Tether, Ripple). Higher values go for platforms supporting smart contracts written in ad-hoc languages (such as Solidity for Ethereum) or general purpose languages (EOS, TRON).


Moving to Performance and Scalability, again the cryptocurrencies that inherited Bitcoins' implementation are at the low tier, given the upper bounds by design on block production rate and transactions per second. However, some technological enhancements on top of existing blockchains, such as the lightning network\footnote{\url{https://lightning.network/}}, may help cope with those limitations -- although raising security concerns as they move transactions off-chain. A better performance goes for highly scalable blockchains as in EOS (2 blocks/1000+ transactions per second) or the DAG in IOTA (requesting a transaction implies validating two others, scaling the network by design).


As for Security, intended as the vulnerability to attacks, Proof-of-work blockchains (Bitcoin and its forks, Ethereum) offer the better theoretical security, only vulnerable to 51\% (or more) attacks. Alternative consensus algorithms, mainly flavors of PoS and DPoS (EOS, Ripple) may introduce certain degree of centralization, rendering the network vulnerable with 33\% malicious nodes, or even less in certain contexts\cite{sankar2017survey,schwartz2014ripple,castro2002practical}. That being said, the technology itself is not enough to prevent the formation of mining pools, which may practically pose a security threat when acting maliciously\cite{gervais2014bitcoin}.


From the Decentralization point of view, we can see that the more centralized blockchains got the lower scores. Mainly, those relying on a company or institution to provide trust to the network, as in Ripple and Tether. However, a certain degree of centralization may be desirable in certain contexts. For example, when certain organizations (government, large companies) want to run and control access to their private blockchains. The higher scores went to fully decentralized blockchains as in Bitcoin or Bitcoin Cash, with no trusted authority nor a hierarchy among the participants: all nodes are equally capable of proposing and validating transactions. 


Finally, moving to Privacy, the blockchain was born as a transparent and publicly auditable ledger. However, in certain cases the participants on a transaction want to remain undisclosed, as in traditional banking schemes. Public decentralized blockchains got the lower scores (Bitcoin, Ethereum and so on) as they allow tracking the wallets involved in transactions, ultimately making possible taint analysis to quantify associations between pairs of addresses as the number of transactions increase\cite{tennant2017improving}. 
The higher scores, in turn, are given to blockchains that offer privacy by design (Ripple, Tether, Tron) by means of a trusted intermediary (the owner of the network) which can then obfuscate the addresses involved in transactions.



\section{Framework Evaluation}
\label{sec:case-study}

This section presents the empirical validation of \textit{ChainMaster}, featuring real-world blockchain applications analysed by means of case-study research \cite{casestudy}. We employed the well-known Architecture Trade-Off Analysis Method (ATAM) \cite{atam} as applied to the cases in question, featuring structured focus-groups with design decision-makers. First we provide an overview of the ATAM method and its application to our case-studies. 

\subsection{The Architecture Trade-off Analysis Method ATAM}
ATAM is a scenario-based architecture analysis method, focused on multiple quality attributes (e.g., performance, security), and aimed at locating and analyzing tradeoffs in a software architecture \cite{kazman1998architecture}. As key aspects of the method we will use scenarios and utility trees. \textit{Scenarios} are short, "use case"-like descriptions that pose a functional situation on the system and demonstrate how a quality attribute is manifested. They aid to get a concrete understanding of the effect of weighting quality attributes. 
\textcolor{red}{Furthermore, scenarios serve to assess in-practice the quality attributes captured in the utility tree.}

 The goal of the\textit{ utility tree }is to represent the overall usefulness (utility) of the system, by decomposing and refining the quality attributes. Thus, the root of the tree is the utility (level 0), and the quality attributes (i.e., the architectural features in our framework) are level 1. The next level refines the attributes into sub-categories, in our case the technological decisions that affect each feature -- e.g., Performance can be refined to Block Production Rate, Consensus Algorithm, Enhancement technology and TPS.
 
 ATAM helps to decompose and refine the business goals, quality attributes and usage scenarios of the system to better understand them, prioritize them, and ultimately use them for assessing and selecting blockchain technologies for a well-thought solution design.



\subsection{Experimental Setup}
The application of ATAM took place in 4 different instances and served two purposes: (1) the empirical evaluation of \textit{ChainMaster} in the context of four case-studies; (2) an example usage methodology of \textit{ChainMaster} in the cases in question. The main goal was to guide practitioners into informed decisions about selecting the best-suited blockchain architecture for their dApps. 

We selected the case studies covering a broad spectrum of configurations, technical decisions and development stages; as well as the possibility of having direct feedback from the developers and decision makers behind each of these cases.
To control for case-study origin, we adopted two industrial case-studies, namely, \emph{Photofied} (insurance technologies) and \emph{ArtID} (artworks market) as well as two case-studies from academia, namely, \emph{Cobra} (problem-solving gamification) and \emph{Genesy} (genomics research). 

For the tradeoffs analysis, we conducted semi-structured interviews with several developers on each case study, a form of interactive discussion often used in exploratory investigations to understand phenomena and seek new insights\cite{weiss1995learning}.
The general structure of the interviews consisted on some background questions aimed at characterizing the sample of practitioners interviewed, and then questions around each architectural feature\footnote{A sample guide for the interview is available at: \url{https://forms.gle/ZJhpafE5dVRcheMa8}}. Then, we gathered the scenarios for the ATAM analysis \footnote{A detail of the scenarios gathered for our-case studies is available at: \url{https://tinyurl.com/y7oor7ta}} by putting together different use cases, system documentation and developers' insights. Finally, we performed several discussion rounds to clarify and gather further details on the answers. 

Then, we constructed the utility trees for each case, providing a quantifiable overview of the most important features and tradeoffs -- see Figure~\ref{fig:utility-tree} in the Appendix.
One can either finalize the assessment with the valuable insights obtained; or further proceed to select/suggest the most suitable blockchain(s). For example, comparing the scores given by the developers to each feature, with the Chainmaster's insights for the different implementations in Table~\ref{tab:qualitative-analysis}, plus the opinionated expert view in Table~\ref{tab:quantitative-scores}. This gives a glimpse on the similarity between the developers needs and the features/scores for each blockchain implementation.

For the sake of presentation, we first introduce the academic case studies and then the industry ones. COBRA\cite{cobra_gray} (Section \ref{sub:cobra}) represents an early stage project (still under design) and illustrates how the framework can help on selecting the right blockchain at design time. Genesy\cite{genesy2019} (Section \ref{sub:genesy}), represents a case in which the prior choice was a non-blockchain distributed ledger; so the framework helps to envision whether the particular requirements of the application call for an innovative solution other than the "traditional" blockchains.
Photofied (Section \ref{sub:photofied}) is a project already in production, which illustrates how to successfully migrate the underlying blockchain platform thanks to the insights gained from the framework application.  Finally, ArtId\cite{artid} (Section \ref{sub:artid}) illustrates the usefulness of the framework in the context of auditing an existing commercial platform and gaining insights on its design and tradeoffs for further decision making, as a potential customer, manager, investor or stakeholder in general. 


\subsection{COBRA: A Framework for Collaborative Problem Solving}
\label{sub:cobra}


COBRA (Coopetitive Optimization Blockchain for Research Advances) is an innovative crowdsourcing concept that leverages the blockchain along with the use of gamification to foster collective intelligence. COBRA is a marketplace of participants (human, machines) contributing with their resources (computational power, skills, creativity, etc.) towards a predefined research goal -- typically solving an algorithmic optimization problem -- in a decentralized and secure fashion \cite{cobra}.

\begin{figure*}[htb]
 \vspace{-.5cm}
\begin{subfigure}{9cm}
	\centering
    \includegraphics[width=0.7\textwidth]{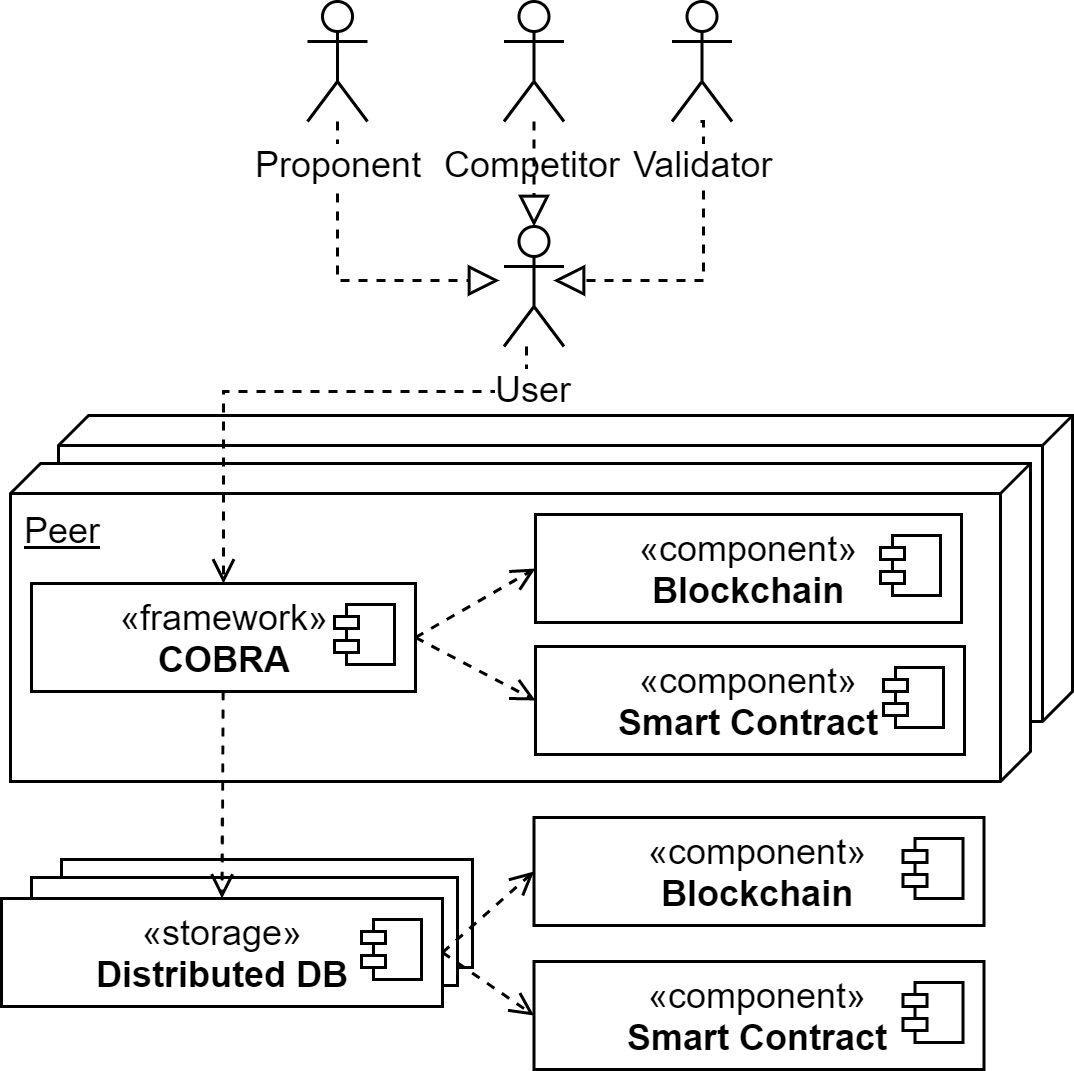}
    \caption{COBRA's architecture.} 
    \vspace{-0.2cm}
	\label{fig:cobra_overview}
\end{subfigure}
~
\begin{subfigure}{9cm}
    \centering
	\includegraphics[width=0.9\textwidth]{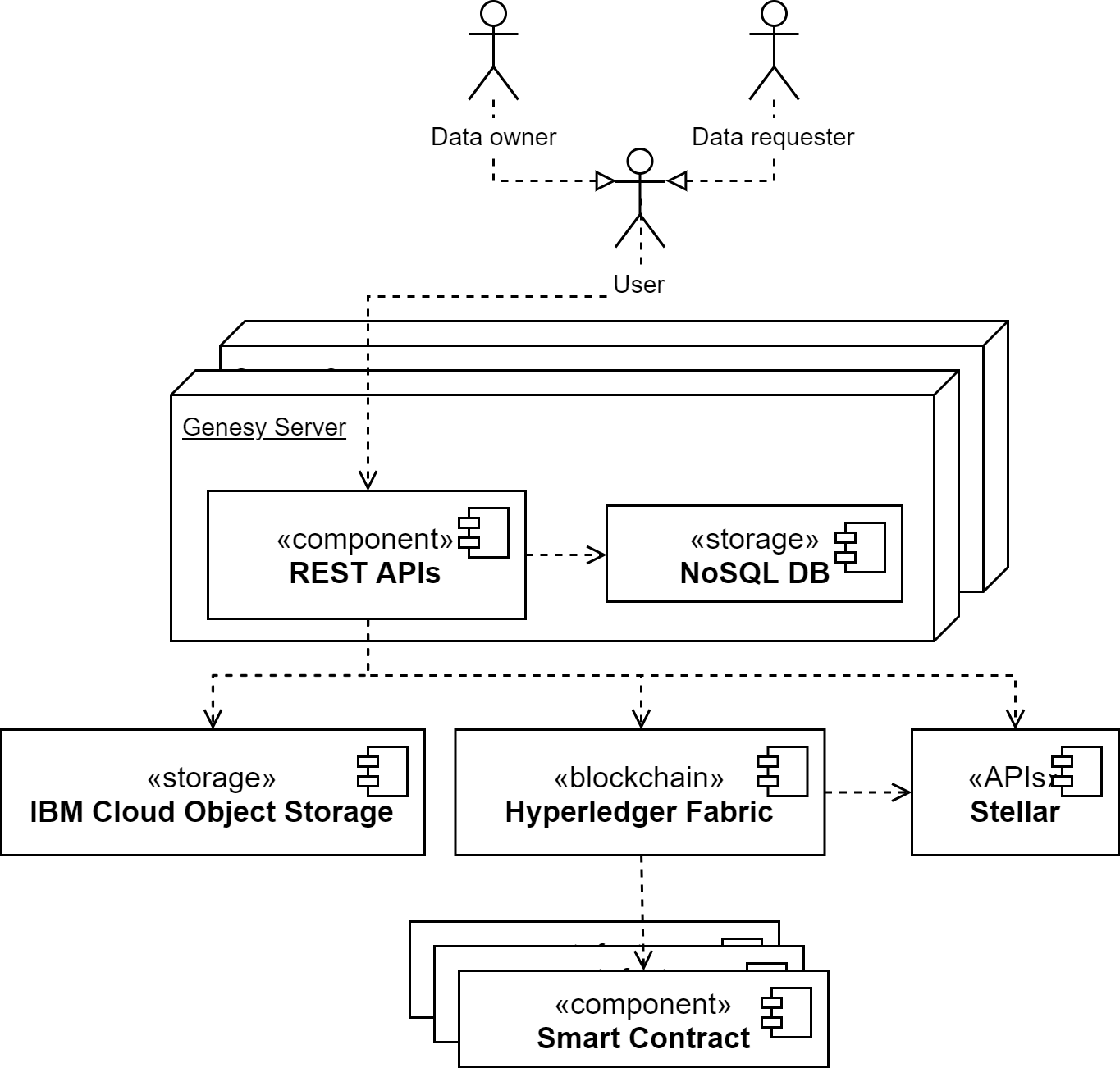}
    \caption{GENESY's architecture.} 
    \vspace{-0.2cm}
    \label{fig:lbl_genesy_model}
\end{subfigure}
\caption{ UML Deployment diagrams for the academic case studies.}
\vspace{-.5cm}
\end{figure*}

Figure \ref{fig:cobra_overview} depicts the architecture of COBRA, where users on the marketplace (i.e., proponents, competitors and validators) can interact with each other and the blockchain through a web interface. 
In particular, a proponent uses the Web application to propose an optimization problem, eventually with a baseline algorithm to solve it and a validation algorithm used to validate solutions' admissibility. The problem is then replicated and stored in the distributed ledger.

Competitors can download the problem and (eventually) the baseline algorithm provided, and try to solve it. 
They can set up their own algorithms, or use existing ones (e.g., exact methods, metaheuristics, etc.), store them into the database and share the obtained solution(s) on the marketplace. At this stage, the validation algorithm is used to verify the solution admissibility before storing it in the database and reward the competitor(s). 

All actions are validated by other peers to guarantee the integrity and the ownership of each operation -- e.g.,  when nodes post, sell, buy or transfer ownership on a solution  (through smart contracts). At any time people can access the marketplace using a Web interface.

\subsubsection{Scenarios}

COBRA is currently at prototyping stage. As developers were not aware of the features of each blockchain platform at design time, it is useful to assess the architecture tradeoffs by means of \textit{ChainMaster} through a set of scenarios.

\noindent{\textbf{Scenario 1}.} Competitors that post valid solutions can be rewarded in three distinct ways: (i)
 by posting the best solution so far; (ii)  by posting any solution as open-source (provided that it is used by others); and (iii) by selling their solution.
 
Thus, it is important to trace the correct order in which the solutions are added to the marketplace to reward the right competitor each time. Hence, \emph{Consistency} and \emph{Performance} become fundamental for the success of the collaboration -- valued \textit{High} by the developers. In particular, \emph{Consistency} ensures that a transaction is securely and correctly appended to the blockchain and guarantee the intellectual property of the solutions. 
Moreover, if the database takes too long to reach a  consistent state, then some competitors -- i.e., those having the database consistent and up to date -- can have undesirable advantages over the others.

\noindent{\textbf{Scenario 2. }}
At any time, stakeholders may want to buy and/or sell solutions, which implies a transfer of ownership. 
Thus, COBRA highly relies on advanced support for complex functionality (i.e., smart contracts) in the underlying blockchain, that triggers when transferring ownership, paying the seller and transferring the solution to the buyer. In that case \emph{Functionality and extensibility} have been considered of high importance while choosing the right blockchain.

\noindent{\textbf{Scenario 3. }}
At any time, any stakeholder should be able to trace down the solutions ownership to their origin. Thus, the developers did not consider \emph{privacy} a key feature. Even more, privacy could be undesirable because COBRA needs the ability to track who posted each solution as well as the solution itself, so if the blockchain obfuscates the origin of the transaction it loses traceability. 
However, one may need to obfuscate a closed-source solution. In that case closed-source solutions are encrypted with the private key of the owner, so in that scenario, partial \emph{privacy} is needed.

\noindent{\textbf{Scenario 4. }}
At any time, new nodes should be able to join the network as to validate/propose transactions, share computational power and query the transaction history, without requesting permission to any centralized trusted authority, thus \emph{decentralization} is a very important feature.

\noindent{\textbf{Scenario 5.}}
Malicious users may want to attack or take advantage of the network by: (i)  Posting sub-optimal solutions to get rewards many times, where the system should pay only for the latest ones and take the rewards back from the others; and (ii) plagiarizing and selling an existing open source solution, where the system could allow the new solution but pay no rewards.




COBRA developers considered that these malicious behaviours do not pose a \textit{Security} issue but rather an exploitation of the \textit{Functionality} -- so the smart contracts should be able to mitigate those misbehaviors. It also increased confidence on the choice of considering low \emph{Privacy}, as the validators should be able to detect and trace who's incurring in that behavior.
In addition, given the niche nature of the application and the research area to which it is intended, the authors are highly confident about the low likelihood of malicious attacks where a single entity would have full control of the majority of the network's mining hash-rate to manipulate it, making \emph{Security} a medium/low feature. Nevertheless, Sybil or other attacks are still possible and the selected platform should make those as impractical as possible without sacrificing other desired features such as Performance.

Finally, a crosscutting feature for all scenarios is \emph{Cost}.  It would be ideal to have minimum fees, fixed costs for validation and variable costs for the rewarding mechanism. Indeed, rewards vary proportionally to the number of existing solutions for the problem at hand as well as to the difference between the value of a posted solution and that of the existing one. 
Instead, validation could simply have fixed costs as it is supposed that validators will spend the same effort to validate the operations (i.e., buy/sell/transfer solutions) on the marketplace.

\subsubsection{Findings and Lessons Learned}

Based on the aforementioned analysis, developers insights, and scenarios, two blockchain implementations emerged to fit COBRA's objectives, namely Tron (TRX) and Etherum (ETH). Although a further analysis should be performed on both options, \textit{ChainMaster} has provided valuable insights. Reasoning on the architectural tradeoffs allowed to conclude that Consistency and Performance are highly important, but those should be achieved in a decentralized fashion -- i.e., without introducing a hierarchy of validator nodes that can reach consistency faster but at the expenses of centralizing the network. Also, protection against malicious behaviours is to be provided at application level -- i.e., it is more a concern of having ample Functionality and Extensibility rather than enforcing Security by design. With that in mind, we were able to further reason on the suitability of existing blockhains by means of ChainMaster and make an informed recommendation in the context of COBRA. Furthermore, this allows for faster prototyping, reducing costs and time to market of the whole application.
\subsection{Genesy: A Blockchain-based Platform for DNA Sequencing}\label{sub:genesy}


\emph{Genesy} is an innovative blockchain platform which acts as an intermediary between the owner of the genomic data and its potential users (i.e., university research centers, private laboratories, hospitals, geneticists, etc.).\cite{genesy2019} It envisions the use of blockchain, cloud computing and artificial intelligence as means to structure a new ecosystem that ensures the user's exclusive property and access to their genomic data, but also the possibility of participating in the benefits of the genomic research. The aim of Genesy is to involve collaboration among users and various organizations to promote a high-level genomic ecosystem, thereby efficiently collecting and managing the large volumes of data produced in DNA sequencing activities.

Figure \ref{fig:lbl_genesy_model} depicts the model under which Genesy operates, where a central authority, namely Genesy, acts as a gatekeeper among peers. Genesy delivers to the customer a kit for DNA collection through vial that preserve dry saliva. This is sent to sequencing structures provided by Genesy itself for analysis. After few weeks the users of the platform receive their results, which are stored in the database where genomic data themselves reside, while access metadata are stored on the blockchain keeping in mind the privacy concerns.
At the same time, the genome data can be sold (with the acknowledgement of the owner) through the same marketplace following the pay-per-use schema, while the ownership of the data still belongs to its original producer.


\subsubsection{Scenarios}

\noindent\textbf{Scenario 1.}
Anonymous data will be shared and later processed with advanced tools for computer science and scientific analysis, taking advantage of batch Big Data technologies. Hence, \emph{Performance and Scalability} is not a must (\textit{Low}). However, \textit{High} or \textit{Very High Privacy} as genomic data should remain anonymous for further users (researchers or companies).

\noindent\textbf{Scenario 2.}
Users will be able to receive and share results through controlled access mechanisms and download information from Genesy locally. Users will analyze and interpret their genomic data in complete autonomy through an ad-hoc application, and a service will regularly notify them of the possibility to acquire new reports applied to their DNA. Access to data from users and professionals/academic or pharmaceutical facilities will take place with identity management guaranteed by private-public key pair systems and cryptographic functions, ensuring the ownership and exclusive access of users genomic data. In summary, Genesy will act as an ecosystem manager and gatekeeper of the information, resulting in a central authority. This way, the \emph{Decentralization} requirement is \textit{very low}, thus steering the decision towards private blockchain implementations that support a high degree of centralization. In addition, as the access mechanism is handled internally by Genesy, \emph{Security} can be \textit{low/medium} in the choice of the desired blockchain.

\noindent\textbf{Scenario 3.}
All transactions between the participants in the marketplace will be carried out in Genesy and regulated by smart contracts. In particular, an ad-hoc currency will allow to purchase and sale data on a linked cryptocurrency network. The Genesy coin can be "exchanged" for a cryptocurrency in the context of two different use cases: (i) when the DNA is sequenced, and (ii) if the user shares his/her DNA data on the platform.

Thus, \emph{Functionality and Extensibility} have a \textit{Very High} impact the choice of the blockchain implementation: both Smart Contracts and interchain communication are required, the former to support the marketplace functionality and the latter to monetize transactions through the sidechain\cite{pilkington2016}.

\noindent\textbf{Scenario 4.}
A user sharing his/her genomic data should be capable of setting his privacy degree. The DNA data will be stored on Genesy's servers and they will be the only ones able to connect the name of the users to their DNA. 
However, users may decide to share it anonymously with external companies, in which case those companies will be able to query them for in-depth analyses, but ownership will remain of the users.
In addition, there is the possibility of attracting the attention of pharmaceutical and research companies that can communicate a special interest for certain DNA. In that case, they will never see the name of the user, but Genesy will notify the users who are left the choice to accept whether or not to contact these organizations for further analysis and then revenue.
Thus, although the \textit{Privacy} requirements, it should be possible, under certain conditions, to access the end user identity. Genesy acts as a gatekeeper, which implies a \textit{Low} \textit{Decentralization}.

In addition to the architecture features considered in the previous scenarios, \emph{Cost} has not been considered to be of importance, while a medium value of importance has been given to \emph{Consistency} while conducting the discussion with developers. Based on all the information gathered, and the insights obtained by means of \textit{ChainMaster} and ATAM notions; the choice of a suitable blockchain was circumscribed to EOS and Ethereum. In particular, the former emerged to satisfy all the requirements gathered from the aforementioned scenarios.

\subsubsection{Findings and Lessons Learned}
At the moment of conducting the analysis, Genesy was at prototyping stage, whilst the technological choices lead to Hyperledger fabric\cite{androulaki2018hyperledger} as the main distributed ledger (for managing genomic data and related functionality), and Stellar\cite{stellar} as a sidechain to back transactions that involved the Genesy cryptocurrency. Hyperledger Fabric is a modular and extensible open-source system for deploying and operating permissioned blockchains. It supports modular consensus protocols and smart contracts written in general-purpose languages (i.e., high \textit{Functionality \& Extensibility}), tailored to particular use cases and trust models. Hyperledger Fabric does not have a currency tied to it and can provide support to any type of dApp without imposing any \textit{Cost} (transactions are theoretically free). Among the non-crypto blockchains, it is one of the most promising and widely adopted to date, although its \textit{Performance} is still to be tested on large scale scenarios~\cite{nasir2018performance}. Indeed, Hyperledger Fabric typical scenarios are private/consortium blockchains with dozens of nodes (as in Genesy). If the number of organizations grows, a layered architecture could be exploited to support the scale while maintaining the performance.

The gap between the suggested and adopted technologies will be discussed in Section \ref{sub:case-studies-discussion}. Both Genesy developers and decision makers agreed at design time that none of the existing blockchains would cope with their particular needs. Specifically, the privacy conditional to the use case scenario (i.e., sometimes the parties should be anonymized but not always), and the Genesy consortium role as a gatekeeper and trust provider among the involved stakeholders. In this scenario, the decision was to implement the particular business logic (including smart contracts, consensus and permissions) using  Hyperledger Fabric as a platform. Being an open source and modular environment, this allows for control on all layers, but at the expense of increased development complexity and costs, and delayed time to market. Through our analysis we demonstrate that existing blockchains could cope with the requirements (e.g., EOS, Ethereum). For the privacy needs, available technology enhancements may prove useful, such as zero-knowledge proofs upon Ethereum\cite{williamson2018aztec}. Additionally, using a "traditional" blockchain suppresses the need for a sidechain backing the cryptocurrencies transactions associated with Genesy. Of course, the final decision is taken by the responsible persons, something definitely beyond the scope of the present analysis.



\subsection{Photofied: A Trusted Images Application for the Insurtech Domain} \label{sub:photofied}


Photofied\footnote{\url{https://photofied.tech}} is a solution for worldwide insurance activity. It certifies digital images in the blockchain for fraud prevention, granting reliability on the status of an insurable risk, both at policy emission
and execution stages.
All images taken with Photofied are certified by means of an ad-hoc protocol, namely Three Way Certification (3WC) featuring blockchain, a P2P distributed file system and digital signature to grant the immutability, perdurability and verifiability of the images. 

\begin{figure*}[htb]
\begin{subfigure}{9cm}
    \includegraphics[width=.85\textwidth]{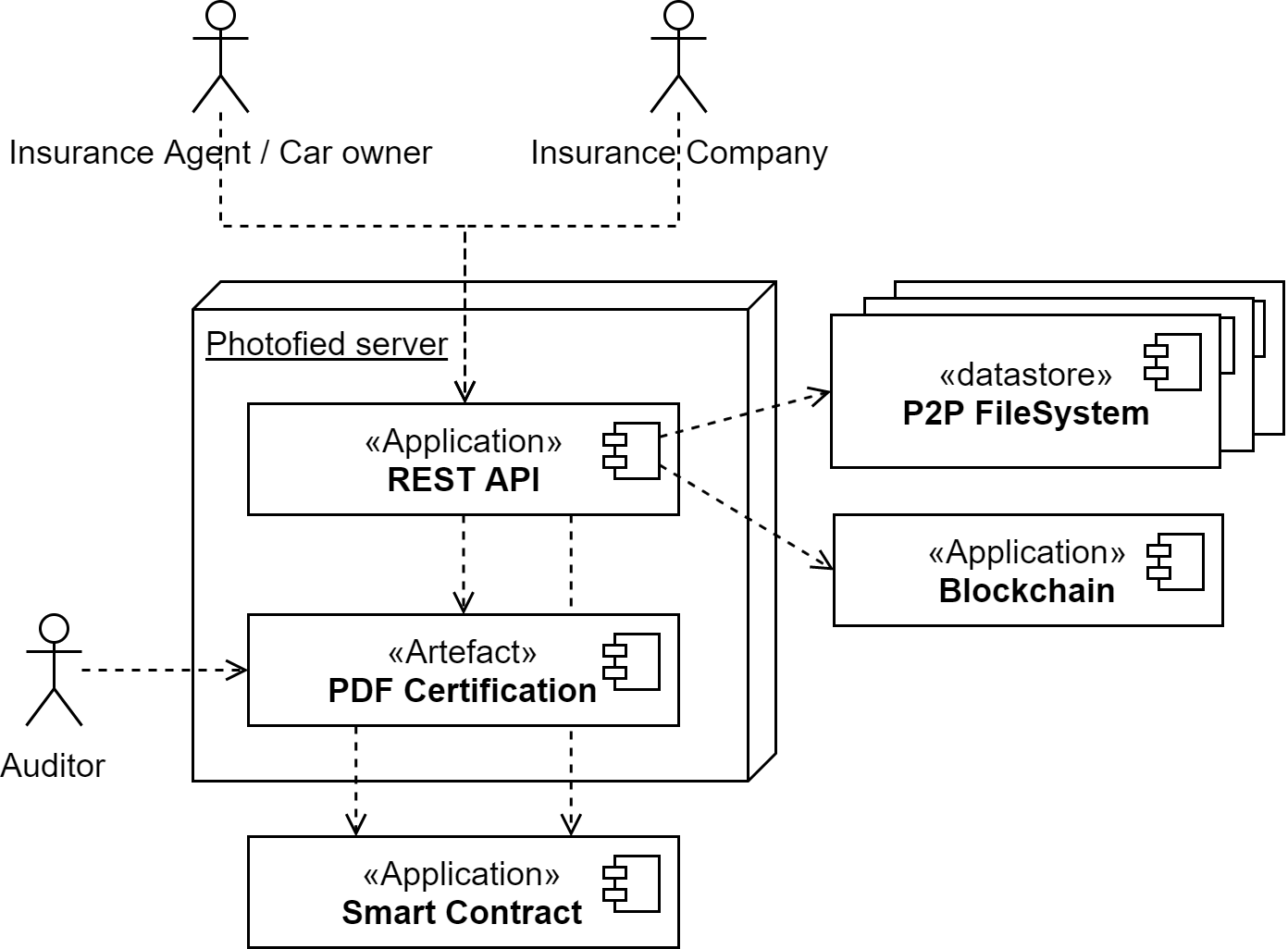}
    \caption{Photofied architecture.}
    \vspace{-0.2cm}
	\label{fig:photofiedArch}
\end{subfigure}
~~~
\begin{subfigure}{9cm}
   \centering
    \includegraphics[width=.75\textwidth]{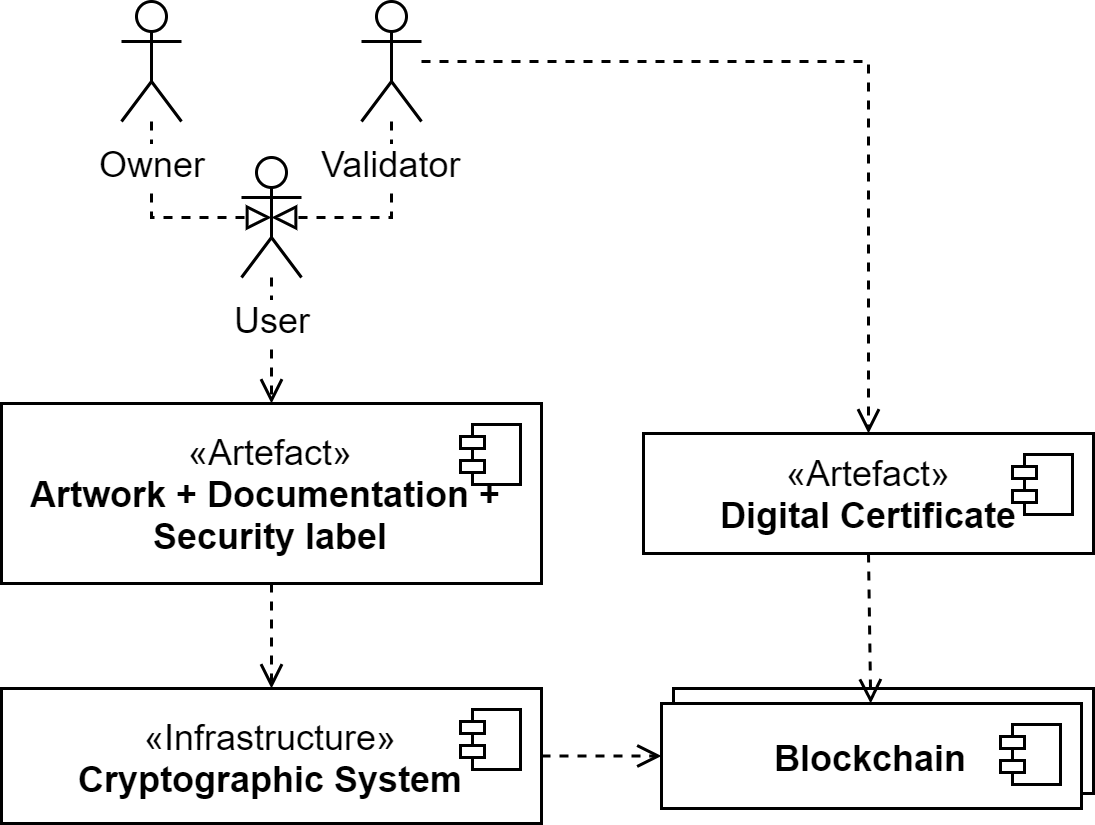}
    \caption{ArtID architecture.}
    \label{fig:artid_overview}
    \vspace{-0.2cm}
\end{subfigure}
\caption{ UML Deployment diagrams for the industry case studies.}
\vspace{-.5cm}
\end{figure*}

Figure \ref{fig:photofiedArch} depicts the architecture of the application, where insurance agents or car owners use the mobile app to send packages (containing images and meta-data to certify) to the Rest API. The server forwards the package to the smart contract and the p2p Filesystem. After that, each certification is printed to PDF, allowing offline audit. At any time, insurance companies can access the certifications using a Web interface.

Similarly to COBRA, at application's design time, the developers were not aware of all the advantages and drawbacks of each blockchain platform. A first selection, purely based on popularity, led first to a Bitcoin-based implementation and then an Ethereum-based one. The former used a na\"ive (\emph{Data hash} $\rightarrow$ \emph{Timestamp}) mapping structure, given the lack of proper smart contracts support in the Bitcoin blockchain. The latest used a smart contract that stored a mapping from each uploaded piece of information to the account from which it was uploaded along with some meta-data.

However, both implementations suffered the drawbacks of the underlying blockchains (See Table \ref{tab:qualitative-analysis} and Table~\ref{tab:quantitative-scores}), requiring expensive transaction fees and relying on congested networks, with a prohibitively low number of transactions per second. In parallel, the number of novel blockchain platforms increased exponentially, paving the way for the adoption of a most suitable one. This can be solved by applying \textit{ChainMaster} to this case. First, developers gained deeper knowledge and understanding of their architecture by isolating the critical features and tradeoffs to be provided by the blockchain through the use of scenarios. 

\subsubsection{Scenarios}

\noindent{\textbf{Scenario 1}}. At a first stage, Photofied is used by insurance agents during the \textit{policy emission process}, certifying the images taken. The images are later audited only if needed, thus there is no need for fast transaction confirmation -- i.e. fast \textit{Consistency} is given a \textit{Low} value as it is not a must. 

\noindent{\textbf{Scenario 2}}. A car owner may want to capture images him/herself and self-generate the insurance policy. The dApp needs to identify who took the images, when, and where. This calls for \textit{High}
\textit{Functionality}. As certifying images independently may harm the performance, they are sent in batches, which reduces the \textit{Performance} requirements (\textit{Medium}) value.

\noindent{\textbf{Scenario 3}}.
End users should be able to query their insurance policies at any time through the app. However, end users are neither supposed to know about blockchain or cryptocurrencies and/or own accounts; nor responsible for paying for the service --- thus, \textit{Cost} should be \textit{low} to attract insurance companies as potential customers. 

\noindent{\textbf{Scenario 4}}.
At audit time, auditors need to access the certified images with certainty that those were not modified in the meantime. That is, \textit{Security} is a concern to maximize, granting the trustability of certified images. Each certification, containing images and meta-data (username, GPS coordinates and mobile device's information), has to be auditable by third parties without using Photofied services/servers (i.e., by querying directly the underlying blockchain), which favours traceability instead of full privacy/obfuscation. As stated for Scenario 2, the target blockchain should provide \textit{High} \textit{Functionality} to support for this process through smart contracts.

\subsubsection{Findings and Lessons Learned}

Guiding this analysis is already a contribution to the developers, since they gain a systematic view to assess the architecture of their \textit{dApp}. 
The last step implies an actual recommendation of the most suitable blockchain. From the tradeoff analysis, the most suitable options resulted EOS and Tron, due to their low cost (no fees), functionality (through smart contracts), good performance and transparency (low privacy), among others. To untie, EOS was selected based on its popularity, as an indicator of more active developers, nodes, available dApps and supporting community. Also, by the time of the decision, the Tron main-net was not yet online and had no near release date (later launched around mid-2018). As a corollary, the current version of Photofied is running using EOS in a collaborate effort with EOS Argentina\footnote{\url{https://www.eosargentina.io/}}, a block producer on the EOS blockchain. 
\subsection{ArtID: A Blockchain Technology for Artwork Certification} \label{sub:artid}


\emph{ArtID} is an application that leverages  blockchain technology and digital certificates to provide certainties to today's risky and low-liquidity art market, which suffers the lack of a controlling authority\cite{artid}.
ArtID responds to the critical issues of that market, specifically to: (i) safeguard the value of an artwork (or collectibles); (ii) reduce the risk associated with the purchase; (iii) increase the liquidity of the artwork in case a sale is needed; and (iv) safeguards the privacy of the owner.

Figure \ref{fig:artid_overview} depicts an overview of ArtID's workflow, where first the artwork is photographed in high definition with the help of an ad-hoc application. The image is digitized and linked to the identity of the artwork and to the documentation of authenticity and provenance, in a single digital record.
In the interest of the owner, each subsequent transaction/action regarding the artwork will be recorded in this file, whose integrity is guaranteed by the use of blockchain technology. 
Once created, the digital identity is authenticated by one or more authorities recognized by the market (e.g., artists, foundations, archives, etc.). A digital certificate will allow to build a new generation of artworks that will differ from the previous ones for the possibility to access all the documentation in real time, through an incorruptible archive. Thus creating the condition for a new, more open, transparent and liquid market.

The Digital Certificate is a zip file signed on the blockchain and contains all the information and documents produced by the owner related to an artwork, verified and checked, both for their authenticity and for their contents by a staff composed of operators and experts who have the title and the ability to verify each individual document both for the authenticity and for the interpretation of the content. The issue of the Digital Certificate involves the generation of a Security Label that once put on the artwork can no longer be removed without evidence of tampering. The sticker is attached to the back of the artwork and allows access to the Digital Certificate through the use of the public key on it or via QR Code accessible through the mobile applications.


\subsubsection{Scenarios}

\noindent\textbf{Scenario 1.}
On ArtID, an artwork's owner can have all the information about it, recorded permanently in digital form, instantly transferable and accessible in a safe and anonymous way, requiring the blockchain to provide \textit{high Performance} and \textit{Privacy} to safeguard the anonymity of the owner and sellers.

\noindent\textbf{Scenario 2.}
ArtID provides a closed-source proprietary process to execute three fundamental operations, namely (i) registration and issue of the identify document of an artwork; (ii) permanent digital authentication; and (iii) 
support the possibility to compare the physical artwork with the high-definition images in the Digital Certificate signed on Blockchain. These three operations require a \textit{High} degree of \textit{Functionality and Extensibility}, allowing to code the smart contracts to do so.

\noindent\textbf{Scenario 3.}
A \textit{marketplace} facilitates the exchange process between artworks and tokens, while protecting buyers and sellers.
Only the artworks equipped with the Digital Certificate can be listed on the marketplace. Those be acquired using the ArtID tokens, which in turn can be obtained in the following ways: (i) by becoming a validation authority; (ii) by becoming a seller (art gallery, collector, academy, etc.) and selling artworks on the ArtID marketplace; and (iii) by purchasing them from exchanges/ArtID. Again this calls for \textit{high} Functionality and particularly the support for creating ad-hoc tokens.

\noindent\textbf{Scenario 4.}
An Escrow Service guarantees maximum protection, withholding the tokens used in exchanges until the buyer, after receiving and evaluating the work, authorize the transaction. All the documents relating to the artworks listed on the Marketplace are already assessed but their relation with the artworks requires the usual appropriate checks. In the absence of
disputes within the set time the tokens are automatically freed. The escrow service is indeed a smart-contract, but the checks should be performed by authorized nodes (validators). This arises the need for supporting certain degree of \textit{centralization}, in which certain nodes can validate the transactions.

\subsubsection{Findings and Lessons Learned}

The current ArtID platform is implemented upon Ethereum. From the analysis of the scenarios and the answers from the semi-structured interview, we concluded that Ethereum is not the most suitable blockchain for the architectural features that should be maximized. In particular, there are requirements from the use case stakeholders for certain validation privileges -- as only museums, art galleries and specialists could validate the authenticity of an artwork/digital certificate. This may imply certain degree of centralization; conflicting with requirements on high Performance and Security (recall the Scalability trilemma discussed in Section~\ref{sub:summary}). Another alternative would be to give some trusted nodes the ability to submit signed transactions -- so the validation is decentralized. Application developers considered this also as a partially centralized solution, as it still introduces some authoritative nodes (which should be able to interact with the blockchain and possess a private key). Besides, security could be tampered because more nodes would be able to create certificates, so more guarantees should be provided over identity.

All in all, the most suitable blockchains arising from \textit{ChainMaster}'s analysis were EOS, Tron and Cardano, taking into account that any of those incur in the aforementioned tradeoff.
Indeed, while EOS seems to fit best the requirements of developers, the adoption of Cardano or Tron would require a slight sacrifice in terms of \emph{performance} or \emph{security} respectively, but still suitable for the problem at hand. However, this should be considered as a suggestion and not a final decision, since in the context of this case study there may be other drivers that are out of our analysis. To name a few, the migration cost, the popularity of the platform (as a technical and market-wise constraint), and the skills of the development team.


\subsection{Findings and Discussion}
\label{sub:case-studies-discussion}
Although some interesting findings were already discussed in the context of each case study, we provide here some takeaways from the overall analysis. Clearly we notice a gap between the chosen blockchains and the analysis arising from applying \textit{ChainMaster}. The reasons are manifold, but we can highlight that both analyses are opinionated. Developers and decision makers may have their own drivers other than the architectural features, e.g., popularity, developers' skills or simply the hype for a given platform. The suggestions arising from our analysis are opinionated as well since they capture the knowledge of the developers participating in the focus group and our own view upon the literature. Nevertheless, we recall once again that the actual contribution are the insights gained from the reflection exercises on the architectures and decisions \textit{per se}. 

Through the utility tree notion we were able to map architecture features to technological decisions and usage scenarios. From this analysis emerged the need to prioritize certain desired features before start making decisions, as multiple features can be affected in different ways by the same technical decisions -- the tradeoffs. For example, both \emph{consistency} and \emph{performance} are influenced by block production rate. Similarly, both \emph{security} and \emph{decentralization} are affected by the ledger implementation (e.g., blockchain, Interledger Protocol, etc.), while \emph{performance}, \emph{decentralization} and \emph{privacy} are affected by the technology. 
Hence, the developers should take care of the priority of their architecture features. For example, one can expect that improving \emph{privacy} on the application can require additional checks in detriment of its \emph{performance}, thus limiting the choice of the block production rate (in case there are no overlap between choices). The analogue for \emph{consistency} and \emph{performance}.
The same could happen for \emph{security} and \emph{decentralization}, where one blockchain could present strong decentralization properties while ignoring security or vice-versa. As stated by the scalability trilemma\cite{buterin2017trilemma}, fixing scalability requires to sacrify one between decentralization and security, and so the choice of the ledger (i.e., the common technical decision). This further confirms the early findings arising from the Grounded-Theory, discussed in Section~\ref{sub:summary}.

In COBRA and Photofied we were able to successfully apply \textit{ChainMaster} either to aid in the migration to a new blockchain or to circumscribe the decision at design time to a handful of platforms. In Genesy and ArtID,  the decision made by the developers was not inline with the outcomes of applying \textit{ChainMaster}. We believe that further rounds of discussion with the developers are necessary, to fully understand the drivers behind their decisions, but also to capture those within a further extension of the framework, as a future work.

\subsection{Limitations of the study}



\label{sub:assessment-tech}

\textcolor{red}{First, the scope of our analysis is limited to blockchains and distributed ledgers implementing cryptocurrencies. Furthermore, 5 out of 11 top implementations are only intended to be cryptocurrencies (Bitcoin and its forks, Ripple, Tether) and as such these do not support advanced dApps.} However, it would be worth considering also other technologies not necessarily tied to cryptocurrencies -- e.g., Hyperledger, which is briefly discussed in Section~\ref{sub:genesy}. As mentioned in Section~\ref{sub:assessment-tech}, cryptocurrencies are the main use case for blockchains nowadays. That being said, the ecosystem of considered solutions also offers variety, encompassing several blockchain alternatives (e.g., based on PoW, PoS, DPoS) and  non-blockchain ones as well (e.g., the DAG in IOTA and the Ripple ledger). This suggests the applicability of our framework to assess other existing and forthcoming blockchains and distributed ledgers.

Currently, only a handful of experiences in real-world dApps exists\footnote{\url{https://dappradar.com/}}, and still a lot of controversy on whether an application requires the use of blockchain~\cite{lo2017evaluating}. The numerous new entrants like Alogrand\cite{algorand}, Hashgraph\cite{hedera}, Stellar, MaidSafe\cite{midsafe}, etc. highlight the limits of existing blockchains as well. Its debatable whether it is far too early for most companies to seriously commit to any particular blockchain except for small-scale, copy-cat scenarios to test the waters. The emergence of frameworks like \textit{ChainMaster} may help to overcome such difficulties.  

Another important concern in this direction is what we call \textit{Volatility}. It refers to the time-span during which the data, stored in the blockchain, is valid, relevant and/or reliable. Stored data may cease to be relevant when a blockchain project stops working or loses confidence. For example, if a blockchain at a particular moment has only a few  nodes, it may lose confidence since the chain is vulnerable. Another scenario is when a blockchain project grinds to a halt: Up to May, 2019 there are around 2160 different cryptocurrencies with probably as many underlying blockchains\footnote{According to CoinMarketCap -- \url{https://coinmarketcap.com/all/views/all/}}. The evolution of the field have demonstrated that the vast majority of these project will not have continuity. Back to the inclusion criteria adopted for this study, being behind a successful cryptocurrency may suggest that such blockchain will last on time, which is clearly desirable when building solutions upon a given platform. For these reasons, volatility can directly impact blockchain-based dApps, and cannot be predicted. However, as our analysis is validated on the top cryptocurrencies, we expect them to be long-term projects and, as such, their features may prove representative enough to analyze other cases and applicability.

Besides, we are aware that relying only on the focus group analysis might result on a possible bias -- e.g., for the case studies the recommendation favours Ethereum, EOS and Tron more than others -- although we provide the insights from this focus group to show the potential applicability of the framework, not to formally validate it.  In addition, experts may have taken the current state of the corresponding networks regarding hash power, decentralization,  price, etc. \textcolor{red}{Time-dependant insights might become stale fast in an ever-changing scenario such as the distributed ledger technologies (especially for the less adopted). Sudden events (e.g. a flaw found in a protocol or a novel attack) may cause sudden changes of opinions. To mitigate this, the framework will feature continuous feedback from experts and community to represent the current state-of-practice.}

Combining the outcomes of the focus group with the case studies in Section~\ref{sec:case-study} provides a broader picture regarding the potential of \textit{ChainMaster}. One should also note that the consensus among trusted experts in the blockchain world is not stable yet, as it is a new technology with surrounding controversy.  A deeper study involving a wider community of developers and more implementations is needed to gather enough data and show how the features influence each other, by e.g., correlation analysis, in order to guide the choice of the right blockchain more objectively.

\section{Conclusions and Future Work} \label{sec:conclusion}

Blockchain is still an emerging technology, thus not a lot of developers are concerned with the principles of Software Engineering applied to blockchain-based systems. 
Nevertheless, the number and variety of existing blockchain implementations continues to increase. 
Moreover, the lack of guidelines and standards on how to design software architectures that include smart contracts as part of the system calls for further attention \cite{xu2017taxonomy,destefanis2018smart}.
Therefore, adopters should focus on selecting the solution that best fits their needs and the requirements of their decentralized applications (dApps) --- rather than developing yet another blockchain from scratch.

In this paper we presented \emph{ChainMaster}, a conceptual framework to aid solution architects, developers, and decision makers to adopt the right blockchain or distributed ledger technology for their problem at hand. The framework exposes the interplay between technological decisions (such as consensus protocols and support for smart contracts) and architectural features (such as cost and decentralization). 

For crafting the framework, we leveraged the knowledge from different sources: first, a literature review comprising both academic  and \textit{grey} literature (industrial products, technical forums/blogs, white papers, etc.), followed by a focus group with experts from industry and academia; plus our own experience using and developing blockchain-based applications. 

We have shown the suitability of \textit{ChainMaster} and provided insights towards its practical application in two ways. First, we applied it to analyze the  blockchain implementations behind the most popular cryptocurrencies, according to their market cap. This shed light regarding the current ecosystem of mainstream blockchain solutions. The assessment was conducted accounting both the technical decisions and the architectural features. Second, we shown how \textit{ChainMaster} can be applied by dApps developers through four real-world case studies, two from industry and two from academia: (i) a decentralized application for collaborative problem solving; (ii) a blockchain-based application for DNA sequencing; (iii) a trusted images application for the insurtech domain; and (iv) an application for artwork certification. The process ultimately led to the selection and/or successful migration to suitable blockchain technologies in some cases; and in others to a deeper understanding of the architectures and their tradeoffs. 
Our future work comprises fine-tuning the framework by engaging yet more experts from the blockchain world. Afterwards we plan to assess the top 50 implementations to have a complete panorama of the existing solutions, beyond the Bitcoin and Ethereum hype.  We are currently developing a series of questions in the form of a wizard, to guide practitioners in the use of the \textit{ChainMaster} framework as support to decision making for selecting the most suitable blockchain for their applications.

\section*{Acknowledgments}
The authors would like to thank to the anonymous reviewers for their valuable feedback. Also, to Fidtech's team and ArtID's team for their helpful insights. 
This research is partially supported by the European Commission grant no. 825480 (H2020), SODALITE, and by the grant no. 825040 (H2020), RADON; finally, by grant ANPCyT PICT-2017-1725.
\bibliography{references}

\pagebreak

\section*{Appendix}
\label{sec:appendix}

\begin{figure}[ht!]
\centering
\includegraphics[width=1\textwidth]{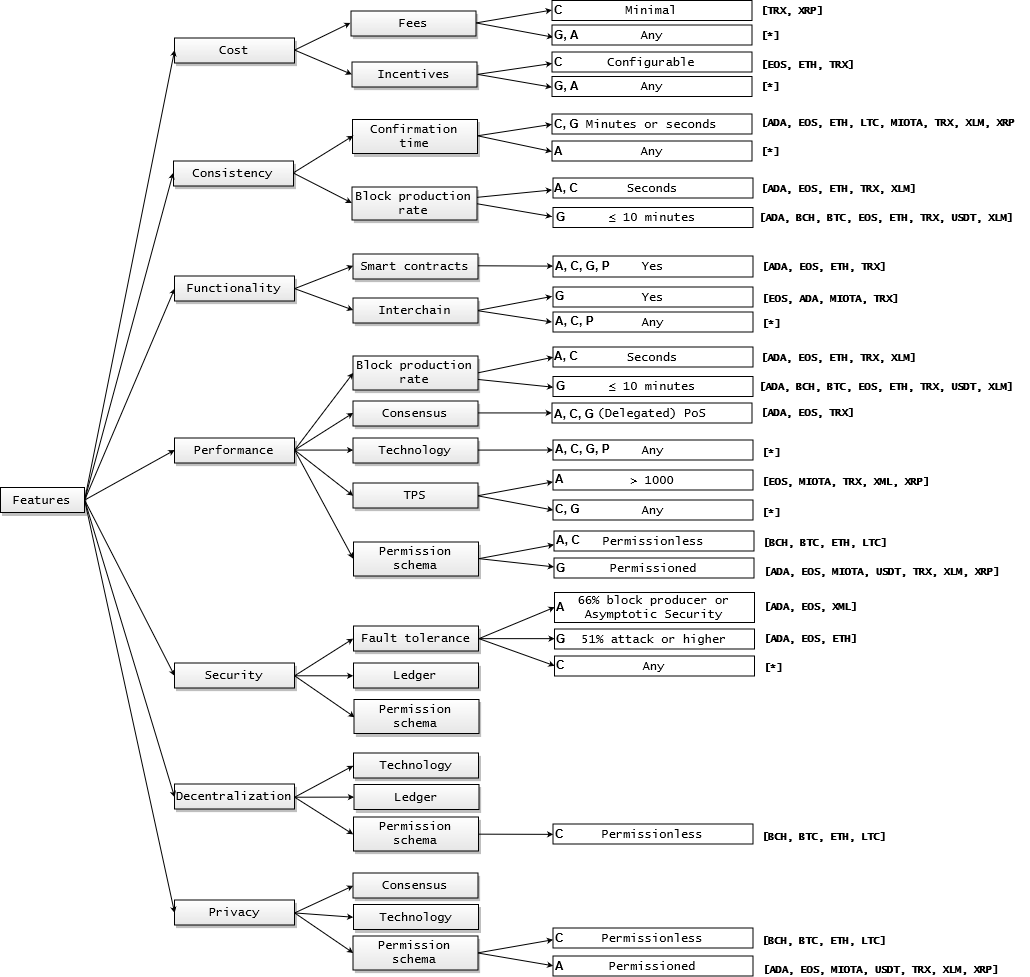}
\caption{Summary utility tree for the case studies. It captures the most important technological decisions behind the architectural features for the four case studies, and the blockchain implementations that may be able to cope with their requirements. Usage scenarios then help to assess such features in practice. \textit{Legenda: A (ArtID), C (Cobra), G (Genesy), and P (Photofied), the four case studies}.} 
\label{fig:utility-tree}
\end{figure}


\end{document}